\documentclass[a4paper,11pt]{article}
\pdfoutput=1 

\usepackage{jheppub}

\usepackage{float}
\usepackage{tabularx}
\usepackage{diagbox}
\usepackage{booktabs}
\usepackage{amsmath}
\usepackage{amsfonts}
\usepackage[rgb]{xcolor}
\usepackage{hyperref}

\usepackage{epsfig}
\usepackage{lineno}
\usepackage{epstopdf}
\newcommand{\be}{\begin{equation}}
\newcommand{\ee}{\end{equation}}
\newcommand{\ba}{\begin{eqnarray}}
\newcommand{\ea}{\end{eqnarray}}

\newcommand{\la}{\label}

\begin{document}
\title{The complex heavy-quark potential in an anisotropic quark-gluon plasma -- Statics and dynamics}

\author[a,b,c]{Lihua Dong,}
\author[a,b]{Yun Guo,}
\author[d]{Ajaharul Islam,} 
\author[e]{Alexander Rothkopf}
\author[d]{and Michael Strickland}

\affiliation[a]{Department of Physics, Guangxi Normal University, Guilin, 541004, China}
\affiliation[b]{Guangxi Key Laboratory of Nuclear Physics and Technology, Guilin, 541004, China}
\affiliation[c]{School of Physics and Astronomy, Sun Yat-Sen University, Zhuhai, 519082, China}
\affiliation[d]{Department of Physics, Kent State University, Kent, OH 44242, United States}
\affiliation[e]{Faculty of Science and Technology, University of Stavanger, Stavanger, NO-4036, Norway}

\emailAdd{donglh6@mail2.sysu.edu.cn, yunguo@mailbox.gxnu.edu.cn, aislam2@kent.edu, alexander.rothkopf@uis.no, mstrick6@kent.edu}

\abstract{
We generalize a complex heavy-quark potential model from an isotropic QCD plasma to an anisotropic one by replacing the Debye mass $m_D$ with an anisotropic screening mass depending on the quark pair alignment with respect to the direction of anisotropy. 
Such an angle-dependent mass is determined by matching the perturbative contributions in the potential model to the exact result obtained in the Hard-Thermal-Loop resummed perturbation theory. An advantage of the resulting potential model is that its angular dependence can be effectively described by using a set of angle-averaged screening masses as proposed in our previous work. Consequently, one could solve a one-dimensional Schr\"odinger equation with a potential model built by changing the anisotropic screening masses into the corresponding angle-averaged ones, and reproduce the full three-dimensional results for the binding energies and decay widths of low-lying quarkonium bound states to very high accuracy. Finally, turning to dynamics, we demonstrate that the one-dimensional effective potential can accurately describe the time evolution of the vacuum overlaps obtained using the full three-dimensional anisotropic potential.  This includes the splitting of different p-wave polarizations.
}

\maketitle

\section{Introduction}
\la{intro}

During the last decades, many efforts have been made to obtain convincing evidence of the existence of a new form of matter, the so-called quark-gluon plasma (QGP), in relativistic heavy-ion experiments. As a sensitive probe to study the hot and dense medium~\cite{Matsui:1986dk,Karsch:1987pv}, the nuclear modification factor $R_{\rm AA}$ of heavy quarkonium states, such as $J/\Psi$ and $\Upsilon$ has been widely discussed in various systems at RHIC and LHC. Having $R_{\rm AA}<1$ indicates that the formation of bound states is suppressed in nucleus-nucleus collisions, relative to that in proton-proton collisions~\cite{Reed:2011fr,CMS:2018zza,ALICE:2018wzm}. It is expected that the successive dissociation of heavy quarkonia, from the weakly bound excited states to the low-lying ones, can serve as a versatile probe of static and dynamic QGP properties (for reviews see e.g.~\cite{Mocsy:2013syh,Rothkopf:2019ipj}).


Due to the fact the heavy-quark relative velocities are small ($v \ll 1)$ studies on quarkonia can be carried out in the nonrelativistic limit.  As a consequence, many in-medium properties, such as in-medium masses and decay rates of the heavy-quark (HQ) bound states can be obtained by solving the quantum mechanical Schr\"odinger equation with a complex HQ potential. Besides a real part that determines the binding energy, the potential also acquires an imaginary part, which is induced by Landau damping of the low-frequency gauge fields together with color singlet-octet transitions. This imaginary part provides information on the decay of a quarkonium state~\cite{Laine:2006ns,Brambilla:2008cx,Beraudo:2007ky,Escobedo:2008sy,Brambilla:2010vq,Brambilla:2011sg,Brambilla:2013dpa} via wavefunction decoherence. It is obvious that establishing a HQ potential, which accurately describes the interactions between the quark and anti-quark, is key to ensuring the success of the Schr\"odinger equation based approach. The HQ potential at short distances can be obtained by employing the hard-thermal-loop (HTL) resummed perturbation theory in the weak-coupling limit, while the commonly used way to study the non-perturbative contributions is to construct phenomenological potential models. Recently, important progress has been made on the measurements of the complex HQ potential using first principle lattice simulations~\cite{Rothkopf:2011db,Bazavov:2014kva,Burnier:2014ssa,Burnier:2015tda,Burnier:2016mxc} and several attempts to develop complex-valued potential models have been put forward for quantitatively capturing the in-medium properties of quarkonia~\cite{Strickland:2011aa,Thakur:2013nia,Krouppa:2015yoa,Burnier:2015nsa,Krouppa:2017jlg,Guo:2018vwy,Lafferty:2019jpr,Islam:2020gdv,Islam:2020bnp}. 

While most of the above applications have focussed on an equilibrium QGP where Fermi-Dirac and Bose-Einstein statistical distributions were used, during the last decade much attention has been focused on exploring quarkonium physics beyond the equilibrium limit by incorporating momentum-space anisotropies generated by longitudinal expansion into the parton distribution function~\cite{Krouppa:2017jlg,Dumitru:2009ni,Margotta:2011ta}. In general, an anisotropic plasma can be either in equilibrium or out of equilibrium. The state of equilibrium, being static and homogeneous, is sometimes anisotropic due to external fields, for example, in a magnetized plasma. Anisotropic states are also common for systems which are out of equilibrium. One example is the QGP generated in the early stages of ultra-relativistic heavy ion collisions. In this case, a momentum-space anisotropy naturally arises due to the different expansion rates in the transverse and longitudinal directions which is a universal feature emerging in both the weak and strong coupling limits. In this work, we will focus on the expansion-induced anisotropies.

Generally speaking, in the high temperature limit one can construct an effective kinetic theory which governs the time evolution of the one-particle quark/gluon distribution functions. An important step towards a realistic phenomenology in heavy-ion collisions can be achieved by adopting the following anisotropic distribution function of Romatschke-Strickland form~\cite{Romatschke:2003ms}  
\be\label{anisodis}
f_{\rm aniso}^{\rm LRF} ({\bf k})\equiv f_{\rm iso}\!\left(\frac{1}{\lambda}\sqrt{{\bf k}^2+\xi ({\bf k}\cdot {\bf n})^2}\right)\, .
\ee
It takes into account the rapid longitudinal expansion of the QGP, and thus represents a tractable way to introduce pressure anisotropies in the local rest frame (LRF) in a kinetic theory approach~\cite{Strickland:2014pga,Berges:2020fwq}.
In eq.~(\ref{anisodis}), $f_{\rm iso}$ is an arbitrary isotropic distribution function and the anisotropic distribution can be obtained by stretching or squeezing $f_{\rm iso}$ along the direction of anisotropy denoted by the unit vector $\bf{n}$. In addition, $\lambda$ is a temperature-like scale which, only in the thermal equilibrium limit, should be understood as the temperature $T$ of the system.\footnote{However, we still call $\lambda$ temperature for simplicity in the rest of the paper.} An adjustable parameter $\xi$ in the range $-1 < \xi < \infty$ is used to quantify the degree of momentum-space anisotropy
\begin{equation}
\xi = \frac{1}{2}\frac{\langle \bf k^{2}_{\perp}\rangle}{\langle k^2_z\rangle}-1\, ,
\end{equation}
where $k_z \equiv \bf k \cdot n$ and $\bf k_{\perp}\equiv \bf k-\bf n(k\cdot n)$ correspond to the particle momenta along and perpendicular to the direction of anisotropy, respectively. By assuming $\bf{n}$ to be parallel to the beam-line direction, $\xi > 0$ corresponds to a contraction of the isotropic distribution in the $\bf{n}$ direction, which is the case relevant to high-energy heavy-ion experiments.

Numerous work has been devoted to investigate new phenomena arising in the presence of a momentum-space anisotropy. The relevant studies cover a very wide range, from collective modes~\cite{Mrowczynski:1994xv,Mrowczynski:1996vh,Randrup:2003cw,Romatschke:2003ms,Carrington:2014bla,Kumar:2017bja} and quarkonium physics~\cite{Strickland:2011aa,Thakur:2013nia,Krouppa:2017jlg,Islam:2020gdv,Islam:2020bnp,Dumitru:2009ni}, to jet energy loss~\cite{Baier:2008js,Roy:2010zg,Carrington:2015xca,Hauksson:2021okc} and transport coefficients~\cite{Srivastava:2015via,Thakur:2017hfc,Rath:2019vvi,Jiang:2020lgw}, as well as the development of an anisotropic hydrodynamics~\cite{Florkowski:2010cf,Strickland:2014pga}. Some of them have already suggested experimentally detectable signals for the evidence of a non-zero anisotropy~\cite{Dumitru:2009ni}. As is universally accepted, almost all of the above mentioned works adopted eq.~(\ref{anisodis}) as a starting point. It should be pointed out that unlike the equilibrium distribution functions, the explicit form of the distribution function in a non-equilibrium setting is not universal. Here, and in many previous works eq.~(\ref{anisodis}) is used without any
justification other than its simplicity, it captures the most important features of an anisotropic QGP which are the emergence of large pressure anisotropies and the appearance of the so-called chromo-Weibel instabilities~\cite{Weibel:1959zz,Mrowczynski:1993qm} 

In the present work, we are interested in the quarkonium physics in an anisotropic QGP where the parton distribution is described by eq.~(\ref{anisodis}). Due to the existence of a preferred direction $\bf{n}$, the spherical symmetry in the HQ potential is explicitly broken. Consequently, an angular dependence emerges in the anisotropic HQ potential denoted by $V (r,\theta,\xi)$ where $\theta$ is the angle of the quark pair alignment $\bf{r}$ with respect to $\bf{n}$.  As one significant part of this work, we aim to construct an accurate model for $V (r,\theta,\xi)$ which includes both the real and imaginary part and can be used as a basic input for quarkonium studies in a non-equilibrium QGP. However, this is challenging because first principle measurements of this quantity are not available at present. One may instead attempt to incorporate momentum-anisotropy effects through a generalization of the well established isotropic potential models with the fewest possible extra assumptions. On the other hand, solving the Schr\"odinger equation with an angle-dependent potential turns into a genuinely three-dimensional (3D) problem which certainly involves increased numerical cost. In practice, to investigate the evolution of the reduced density matrix of in-medium quarkonium, one must solve a 3D stochastic Schr\"odinger equation or corresponding Lindblad equation~\cite{Brambilla:2017zei,Miura:2019ssi,Alund:2020ctu,Brambilla:2020qwo,Brambilla:2021wkt,Omar:2021kra}, which is as of yet numerically prohibitive. Whether it is possible to have an effectively isotropic potential model that reproduces the full 3D results on the binding energies and decay widths of different bound states turns out to be very crucial when one attempts to include the effect of momentum anisotropies on quarkonium evolution using real-time solution of the Schr\"odinger equation. Previous work has assessed this conjecture for a real-valued potential model~\cite{Dong:2021gnb}. However, to do it properly one must demonstrate that the same logic can be applied to the imaginary part as well. 

Based on the above discussions, in this paper we develop a complex HQ potential model for non-zero momentum-space anisotropy which can be further reduced to an effectively isotropic model to simplify the numerical treatment. The rest of the paper is organized as follows. In sec.~\ref{isomod}, we consider an isotropic HQ potential model proposed in previous work, which can be taken as an effective form for the corresponding anisotropic potential. In sec.~\ref{apot}, we determine the angle-dependent anisotropic screening masses 
through which the non-perturbative string contributions in the anisotropic potential model are obtained by implementing the so-called ``minimal" extension. With the potential model in hand, we study quarkonium states with extremely large quark masses by means of perturbation theory in quantum mechanics in sec.~\ref{exhq}. 
The binding energies and decay widths of charmonia and bottomonia are evaluated in sec.~\ref{cb}, where a focus is put on the effects of the momentum anisotropies. Furthermore, with the angle-averaged effective screening masses, we also demonstrate that the resulting 1D effective potential model can reproduce the full 3D results for low-lying quarkonium bound state eigenenergies to high accuracy. In sec.~\ref{realtime}, we consider the real-time solution of the Schr\"odinger equation and investigate the time evolution of the vacuum overlaps for bottomonium states, comparing results obtained using both the 3D anisotropic potential model and the 1D effective potential model. A brief summary and outlook is given in sec.~\ref{con}. In addition, we discuss the uncertainties in the energy splitting of the p-wave states induced by the model dependence in app.~\ref{modde}. In app.~\ref{appa}, the numerical results of the complex eigenenergies obtained by solving the Schr\"odinger equation are listed for different quarkonium states. Finally, the time evolution of the vacuum overlaps for charmonium states are provided in app.~\ref{app:charmonium}.

\section{The complex HQ potential model in an isotropic QCD plasma}
\la{isomod}

As proposed in ref.~\cite{Guo:2018vwy}, the complex HQ potential model in an isotropic QCD plasma is defined by a Fourier transform of the real time resummed gluon propagator in the static limit. Such a gluon propagator includes both a perturbative contribution, which is calculable in the HTL resummed perturbation theory, and a non-perturbative string contribution originating from the dimension two gluon condensate. Explicitly, the potential model can be formulated as
\be\la{revdef}
V (\lambda,r)=-g^2 C_F \int \frac{d^3 {\bf p}}{(2\pi)^3} (e^{i {\bf p} \cdot {\bf r}}-1)D^{00}(p_0=0,{\bf p}, \lambda)\, .
\ee
For the perturbative contribution, besides a Debye screened potential as its real part, it also possesses an imaginary part which determines the decay width of a quarkonium state~\cite{Laine:2006ns}. The results are given by
\be\la{revdefpt}
{\rm Re}\,V_{\rm pt} (\lambda,r)=-g^2 C_F \int \frac{d^3 {\bf p}}{(2\pi)^3} (e^{i {\bf p} \cdot {\bf r}}-1)\left(\frac{1}{p^2+m_D^2}-\frac{1}{p^2}\right)\equiv \alpha m_D ({\cal I}_1({\hat r})-1)\, ,
\ee
\be\la{imvdefpt}
{\rm Im}\,V_{\rm pt} (\lambda,r)=-g^2 C_F \int \frac{d^3 {\bf p}}{(2\pi)^3} (e^{i {\bf p} \cdot {\bf r}}-1)  \frac{-\pi \lambda m_D^2}{p(p^2+m_D^2)^2}\equiv \alpha \lambda ({\cal I}_2({\hat r})-1)\, ,
\ee
where ${\cal I}_1({\hat r})$ and ${\cal I}_2({\hat r})$ are given by the following integrals
\ba\la{int1}
{\cal I}_1({\hat r})&=&4\pi \int \frac{d^3 {\hat {\bf p}}}{(2\pi)^3} e^{i {\hat {\bf p}} \cdot {\hat {\bf r}}}\frac{1}{{\hat p}^2({\hat p}^2+1)}=\frac{1-e^{-{\hat r}}}{{\hat r}}\, ,\nonumber \\
{\cal I}_2({\hat r})&=& 4\pi^2\int \frac{d^3 {\hat {\bf p}}}{(2\pi)^3} e^{i {\hat {\bf p}} \cdot {\hat {\bf r}}}\frac{1}{{\hat p}({\hat p}^2+1)^2}= \phi_2({\hat r})\, ,
\ea
with
\begin{equation}
\phi_n({\hat r}) = 2 \int_0^\infty dz \frac{\mathrm{sin}(z {\hat r})}{z {\hat r}}\frac{z}{(z^2 +1)^n}\, .
\end{equation}
Here, we define ${\hat {\bf p}}\equiv{\bf p}/m_D$ and $ {\hat {\bf r}}\equiv{\bf r}m_D$. In addition, the strong coupling is given by $\alpha=g^2 C_F/(4\pi)$. Notice that we have subtracted a term $1/p^2$ in eq.~(\ref{revdefpt}) which makes the $r$-independent part finite and shifts the $r$-dependent part by a Coulombic term $-\alpha/r$.\footnote{The $r$-dependent part is related to the factor $e^{i\bf{p}\cdot\bf{r}}$ in the Fourier transform. The contribution associated with the factor $1$ in the Fourier transform results in an $r$-independent part.}

The perturbative HQ potential is only applicable to bound states with extremely large quark mass so that the separation $r$ between the quark and antiquark is very small. For quarkonium states that have been extensively studied in heavy-ion experiments, such as charmonia and bottomonia, the string contribution in the resummed gluon propagator plays an important role in the determination of their in-medium properties. Its Fourier transform determines the non-perturbative contributions to the HQ potential. Explicitly, we have
\be\la{revdefnpt}
{\rm Re}\,V_{\rm npt} (\lambda,r)=-g^2  C_F m_G^2\int \frac{d^3 {\bf p}}{(2\pi)^3} (e^{i {\bf p} \cdot {\bf r}}-1)\frac{p^2+5m_D^2}{(p^2+m_D^2)^3}\equiv -  \frac{2\sigma}{m_D} ({\cal I}_3({\hat r})-1)\, ,
\ee

\be\la{imvdefnpt}
{\rm Im}\,V_{\rm npt} (\lambda,r)=-g^2 C_F m_G^2 \int \frac{d^3 {\bf p}}{(2\pi)^3} (e^{i {\bf p} \cdot {\bf r}}-1)  \frac{4 \pi \lambda m_D^2(p^2-2m_D^2)}{p(p^2+m_D^2)^4}\equiv  \frac{4\sigma \lambda}{m_D^2}  ({\cal I}_4({\hat r})-1)\, ,
\ee
where $\sigma=\alpha m_G^2/2$ with $m_G^2$ being a dimensionful constant related to the dimension two gluon condensate~\cite{Guo:2018vwy} and
\ba\la{int2}
{\cal I}_3({\hat r})&=&4\pi \int \frac{d^3 {\hat {\bf p}}}{(2\pi)^3} e^{i {\hat {\bf p}} \cdot {\hat {\bf r}}}\frac{{\hat p}^2+5}{({\hat p}^2+1)^3}= (1+{\hat r}/2)e^{-{\hat r}}\, ,\nonumber \\
{\cal I}_4({\hat r})&=&8\pi^2\int \frac{d^3 {\hat {\bf p}}}{(2\pi)^3} e^{i {\hat {\bf p}} \cdot {\hat {\bf r}}}\frac{2-{\hat p}^2}{{\hat p}({\hat p}^2+1)^4}=-2\phi_3({\hat r})+6\phi_4({\hat r})\, .
\ea

In the potential model as discussed above, the ${\hat r}$-dependence appears only in the four dimensionless functions ${\cal I}_i({\hat r})$ with $i=1,2,3$ and $4$, all of which vanish when $r$ is infinitely large. Therefore, the asymptotic behavior at $r\rightarrow \infty$ is entirely determined by the $r$-independent part in the Fourier transform, namely,
\be\la{vinf}
V (\lambda,r\rightarrow\infty)=g^2 C_F \int \frac{d^3 {\bf p}}{(2\pi)^3} D^{00}(p_0=0,{\bf p}, \lambda)=\Big(-\alpha m_D+ \frac{2\sigma}{m_D}\Big)- i \Big(\alpha \lambda+ \frac{4\sigma \lambda}{m_D^2}\Big)\, .
\ee
On the other hand,  the four dimensionless functions become identical at the origin and ${\cal I}_i({\hat r} \rightarrow 0)=1$ leads to a vacuum Coulomb potential with a vanishing imaginary part when $r\rightarrow 0$. This is to be expected since no medium effect can be probed in this limit.

\section{The anisotropic screening mass and HQ potential model in a medium with small anisotropy}\la{apot}

Exhibiting a relatively simple form, the potential model discussed above has been shown to agree well with the simulation data in Lattice QCD~\cite{Guo:2018vwy}. Obviously, generalizing such a model to an anisotropic medium is an important step towards shedding light on the physics of quarkonium beyond the equilibrium approximation. In fact, the anisotropic HQ potential has been calculated within the HTL resummed perturbation theory~\cite{Dumitru:2007hy,Burnier:2009yu,Dumitru:2009fy}. The resulting real part takes a much more complicated form as compared to the Debye screened potential in equilibrium and an analytical expression becomes available only for small anisotropies. On the other hand, the imaginary part is well defined near the equilibrium where $\xi <1$, while its determination for arbitrary $\xi$ is still an open question due to the presence of a pinch singularity in the static gluon propagator~\cite{Nopoush:2017zbu,Hauksson:2020wsm}. 

Let us focus on small anisotropies. Although the perturbative HQ potential $V_{\rm pt}(r, \theta, \xi)$ can be fully studied in a model-independent way, it appears quite useful to use an effective form to describe $V_{\rm pt}(r, \theta, \xi)$. Such an effective form is obtained by replacing the Debye masses in eqs.~(\ref{revdefpt}) and (\ref{imvdefpt}) with the corresponding anisotropic screening masses, which is formally analogous to its isotropic counterpart.  The advantage of doing so lies in two aspects. First, the rather complicated expression of $V_{\rm pt}(r, \theta, \xi)$, as derived in the perturbation theory, can be significantly simplified. More importantly, the introduced anisotropic screening masses which contain all the effects induced by momentum anisotropies, could provide key information to model the non-perturbative HQ potential in the anisotropic QGP. 

The first attempt to develop an anisotropic potential model was carried out in ref.~\cite{Dumitru:2009ni}, where a basic assumption was put forward that the very same screening scale as appears in the Debye screened contributions also shows up in the non-perturbative string contributions. Thus, the ``minimal" extension of an isotropic potential model to non-zero anisotropy consists of replacing the Debye mass $m_D$, in both the perturbative and non-perturbative part, by the anisotropic screening masses extracted from the effective form of $V_{\rm pt}(r, \theta, \xi)$. Although previous works limited their considerations to the real part of the HQ potential, the same idea could also be applied to ${\rm Im}\,V(r,\theta, \xi)$. 
In the following, with the derivation of the anisotropic screening masses, we develop a complex HQ potential model under the aforementioned basic assumption. Because of the ill-defined ${\rm Im}\,V_{\rm pt}(r, \theta, \xi)$ at large anisotropies, extracting an anisotropic screening mass for arbitrary $\xi$ turns out to be beyond the scope of this study and we only concentrate on the small $\xi$ region in the current work.

\subsection{The real part of the HQ potential model for small anisotropy $\xi$ }\la{real}

In a medium with small anisotropy, up to linear order in $\xi$, the real part of the HTL resummed gluon propagator reads~\cite{Dumitru:2009fy}
\be
{\rm Re}\,D^{00} (p_0=0,{\bf p,\xi})=\frac{1}{p^2+m_D^2}+\xi m_D^2\frac{\frac{2}{3}- ({\bf p}\cdot{\bf n})^2/p^2}{(p^2+m_D^2)^2}\,.
\ee
According to eq.~(\ref{revdef}), the perturbative HQ potential has the following analytical form
\be\la{vrexa}
{\rm Re}\,V_{\rm pt}(r, \theta, \xi)=-\alpha \frac{e^{-{\hat r}}}{r}\big[1-\xi f_0({\hat{r}})-\xi f_1({\hat{r}}) \cos (2\theta)\big]-\alpha m_D(1-\xi/6)\, ,
\ee
with
\ba
f_0({\hat{r}})&=&\frac{6(1-e^{{\hat{r}}})+{\hat{r}}[6-{\hat{r}}({\hat{r}}-3)]}{12 {\hat{r}}^2}=-\frac{{\hat{r}}}{6}-\frac{ {\hat{r}}^2}{48}+\cdots\, , \nonumber \\
f_1({\hat{r}})&=&\frac{6(1-e^{{\hat{r}}})+{\hat{r}}[6+{\hat{r}}({\hat{r}}+3)]}{4 {\hat{r}}^2}=- \frac{{\hat{r}}^2}{16}+\cdots\, .
\ea

As $r\rightarrow0$, the above equation reduces to the vacuum Coulomb potential as expected. On the other hand, at infinitely large $r$, ${\rm Re}\,V_{\rm pt} (r\rightarrow\infty, \xi)$ is given by $-\alpha m_D(1-\xi/6)$ which originates from the $r$-independent part in the Fourier transform. Comparing with the corresponding isotropic result, we find that the effect of small anisotropies amounts to a modification of the Debye mass, {\em i.e.}, $m_D\rightarrow m_D(1-\xi/6)$. Given the asymptotic behavior, an effective form for the anisotropic HQ potential based on eq.~(\ref{revdefpt}) can be formulated as
\be\la{vpteff}
{\rm Re}\,V_{\rm pt}(r,\theta, \xi) =\alpha m_D(1-\xi/6) ({\cal I}_1(r {\tilde m}_D(\lambda,\xi,\theta))-1)\, ,
\ee
where the $\theta$-dependence only appears in the dimensionless function ${\cal I}_1({\hat r})$ via the replacement $m_D\rightarrow {\tilde m}_D(\lambda,\xi,\theta)$.  The above effective form has the desired asymptotic behaviors since ${\cal I}_1(r \rightarrow \infty)=0$ and ${\cal I}_1(r \rightarrow 0)=1$. 

The effective form eq.~(\ref{vpteff}) satisfies the requirements in the limiting cases where $r \rightarrow 0$ and $r \rightarrow \infty$. At a finite separation distance between the quark and anti-quark, it is the explicit form of the anisotropic screening mass ${\tilde m}_D(\lambda,\xi,\theta)$ which must be determined to reproduce the exact result given in eq.~(\ref{vrexa}).

In particular, as a unique feature arising in an anisotropic medium, the energy splitting of quarkonium states with non-zero angular momentum needs to be realized through a proper $\theta$-dependence in the screening mass. The determination of ${\tilde m}_D(\lambda,\xi,\theta)$ can be carried out by matching the effective form of ${\rm Re}\,V_{\rm pt}(r,\theta, \xi) $ to the exact result,\footnote{A Coulombic term $-\alpha/r$ should be added to the effective form when matching it to the exact result.} then we arrive at
\ba\la{reratio}
\frac{{\tilde m}_D}{m_D}&=&1+\xi \Big(-\frac{1}{6}+ \frac{6(1-e^{ {\hat r}})+ {\hat r}(6+3 {\hat r}+ {\hat r}^2)}{12  {\hat r}^2 (1+ {\hat r}-e^{ {\hat r}})}(1+3\cos(2\theta)) \Big) + {\cal O}(\xi^2)\, \nonumber \\
&=&1+\xi \Big(-\frac{1}{8}(1-\cos(2\theta))-\frac{1}{180}(1+3\cos(2\theta)){\hat r} \Big)+ {\cal O}(\xi\hat r^2, \xi^2)\, .
\ea
To obtain the above equation, we implicitly assumed that in the presence of a small anisotropy, the induced modification of the Debye mass $m_D$ is also small, so that the difference $({\tilde m}_D-m_D)$ is linearly proportional to $\xi$. Consequently, contributions beyond linear order in $\xi$ in eq.~(\ref{vpteff}) have been dropped. Given the fact that HTL resummed perturbation theory applies when ${\hat r}\le 1$, we also Taylor expand the result with respect to ${\hat r}$ which is given by the second line in eq.~(\ref{reratio}). 

\begin{figure}
\centering
\includegraphics[width=0.45\textwidth]{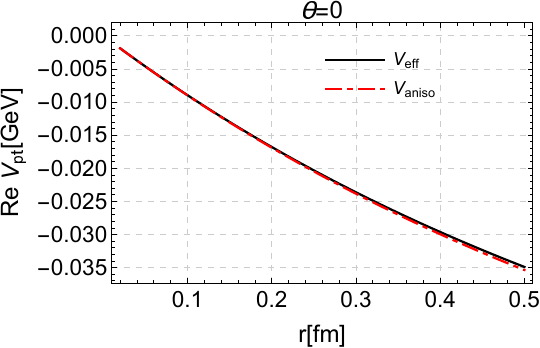}
\includegraphics[width=0.45\textwidth]{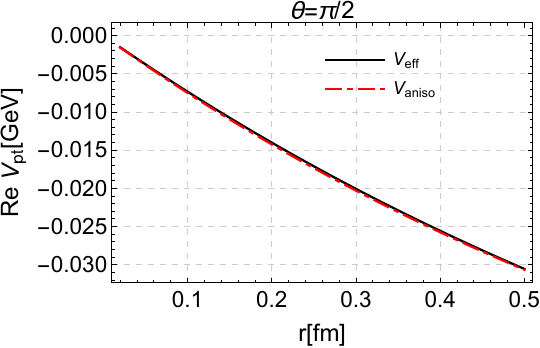}
\vspace*{-0.2cm}
\caption{\label{rbnew}
The real part of the perturbative heavy-quark potential as a function of $r$ evaluated at $\xi=0.7$ for $\theta=0$ (left plot) and $\theta=\pi/2$ (right plot). The solid curves correspond to the exact result while the dash-dotted curves correspond to the effective form eq.~(\ref{vpteff}) with ${\tilde m}_D/m_D = 1-\xi(1-\cos(2\theta))/8 $. We take $\alpha=0.272$ and $m_D=0.4\,{\rm GeV}$. In this figure, a Coulombic term $-\alpha/r$ has been subtracted, therefore, all the curves vanish at the origin.}
\end{figure}

\begin{figure}
\centering
\includegraphics[width=0.45\textwidth]{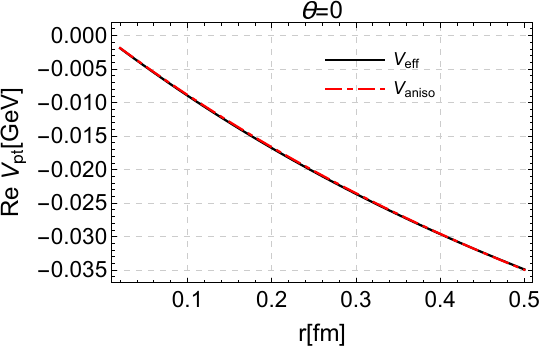}
\includegraphics[width=0.45\textwidth]{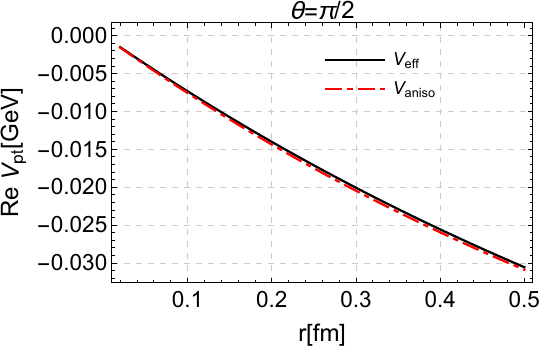}
\vspace*{-0.2cm}
\caption{\label{rb}
The real part of the perturbative heavy-quark potential as a function of $r$ evaluated at $\xi=0.7$ for $\theta=0$ (left plot) and $\theta=\pi/2$ (right plot). The solid curves correspond to the exact result while the dash-dotted curves correspond to the effective form eq.~(\ref{vpteff}) with ${\tilde m}_D/m_D \approx 1-\xi(0.131-0.108 \cos (2\theta))$. We take $\alpha=0.272$ and $m_D=0.4\,{\rm GeV}$. In this figure, a Coulombic term $-\alpha/r$ has been subtracted, therefore, all the curves vanish at the origin.}
\end{figure}

As a function of ${\hat r}$, both eq.~(\ref{vrexa}) and eq.~(\ref{vpteff}) reduce to the vacuum Coulomb potential at leading order in ${\hat r}$. Consequently, eq.~(\ref{vpteff}) becomes identical to the exact result of ${\rm Re}\,V_{\rm pt}(r,\theta, \xi)$ independent on the explicit form of ${\tilde m}_D(\lambda,\xi,\theta)$. The next-to-leading order contribution is linearly proportional to ${\hat r}$ and the matching between the two equations at this order leads to 
\be\la{aeffre0}
{\tilde m}_D/m_D=1-\xi(1-\cos(2\theta))/8=1-\xi(0.125-0.125 \cos (2\theta))\, ,\quad {\rm for}\quad {\hat r}\ll 1\,.
\ee
On the other hand, evaluating eq.~(\ref{reratio}) at ${\hat r}=1$ which can be taken as an upper limit where the perturbation theory works, we obtain 
\be\la{aeffre}
{\tilde m}_D/m_D =1-\xi \frac{6- 2 e - (9 e-24) \cos (2\theta)}{6 e-12}\approx 1-\xi(0.131-0.108 \cos (2\theta))\,  ,\quad {\rm for}\quad {\hat r}= 1\,.
\ee
In fact, eq.~(\ref{reratio}) clearly shows that up to linear order in $\xi$, the exact matching unavoidably requires a ${\hat r}$-dependence introduced in the anisotropic screening mass.  However, it turns out that the determination of ${\tilde m}_D(\lambda,\xi,\theta)$ only weakly depends on the specific values of ${\hat r}$. More importantly, such a ${\hat r}$-dependence is not necessary for phenomenological purposes because eq.~(\ref{vpteff}) is only considered as an approximation to the exact result.  As shown in figs.~\ref{rbnew} and \ref{rb}, a rather good agreement can be achieved by using either eq.~(\ref{aeffre0}) or eq.~(\ref{aeffre}) for ${\tilde m}_D(\lambda,\xi,\theta)$. 

Given the isotropic ${\rm Re}\,V_{\rm pt}(\lambda, r)$ in eq.~(\ref{revdefpt}), the above generalization to the anisotropic QGP involves modifications on the Debye mass. It's worth noting that $m_D$'s in the dimensionless function ${\cal I}_1({\hat r})$ are treated differently from those associated with the $r$-independent part in the Fourier transform because the $\theta$-dependence is only imposed on the former. This is actually consistent with the perturbative HQ potential in eq.~(\ref{vrexa}) where the angle disappears as $r\rightarrow \infty$.

Furthermore, we can generalize the isotropic ${\rm Re}\,V_{\rm npt}(\lambda, r)$ to non-zero $\xi$ by following the idea of ``minimal" extension. According to eq.~(\ref{revdefnpt}), ${\rm Re}\,V_{\rm npt}(\lambda, r)$ reduces to $-2\sigma/m_D$ at asymptotically large $r$ where, in the presence of a small anisotropy, a $\xi$-dependent screening scale $m_D(1-\xi/6)$ needs to be employed, namely $\mathrm{Re}\,V_{\rm npt} (r\rightarrow \infty, \xi)=-2\sigma/(m_D(1-\xi/6))$. For finite separation $r$, a $\theta$-dependence emerges and the anisotropic screening mass ${\tilde m}_D(\lambda,\xi,\theta)$ as determined above needs to be used to replace the Debye mass $m_D$ in the dimensionless function ${\cal I}_3({\hat r})$. Thus, we obtain the following model for the real part of the anisotropic HQ potential
\ba
\label{rV1}
\mathrm{Re}\,V (r, \theta, \xi) &=& - \frac{\alpha}{r} + \alpha m_D(1-\xi/6) ({\cal I}_1(r {\tilde m}_D(\lambda,\xi,\theta))-1)- 2\sigma \frac{{\cal I}_3(r {\tilde m}_D(\lambda,\xi,\theta))-1}{m_D(1-\xi/6)} \nonumber \\
&=&- \frac{\alpha}{r}+\alpha m_D(1-\xi/6)\frac{1-e^{-r {\tilde m}_D}}{r {\tilde m}_D}-\alpha m_D(1-\xi/6)\nonumber \\
&-&\frac{2\sigma}{m_D(1-\xi/6)} e^{-r {\tilde m}_D} (1+\frac{r {\tilde m}_D}{2})+\frac{2\sigma}{m_D(1-\xi/6)}\, .
\ea
When solving the Schr{\"o}dinger equation for charmonia and bottomonia, we also include a relativistic correction in the potential model which is given by $-0.8 \sigma /(m_{b/c}^2 r)$ \cite{Dumitru:2009ni}. The masses of the charm and bottom quarks are taken to be $m_c=1.3\, {\rm GeV}$ and $m_b=4.7\, {\rm GeV}$, respectively.

As $r\rightarrow 0$, the vacuum Coulomb potential can be reproduced from eq.~(\ref{rV1}) which indicates a vanishing non-perturbative contribution in this limit. Indeed, this is guaranteed by construction since ${\cal I}_3(r \rightarrow 0)=1$. On the other hand, at infinitely large $r$, we get
\be
\label{rVinf}
\mathrm{Re}\,V (r\rightarrow \infty, \xi) = -\alpha m_D(1-\xi/6) +\frac{2\sigma}{m_D(1-\xi/6)}\, .
\ee
This result suggests that either increasing $\xi$ or decreasing $m_D$ makes the value of $\mathrm{Re}\,V (r\rightarrow \infty, \xi) $ larger.

\subsection{The imaginary part of the HQ potential model for small anisotropy $\xi$ }\la{sec:im}

For small anisotropies, the imaginary part of the HTL resummed gluon propagator is given by
\be
{\rm Im}\,D^{00}(p_0=0,{\bf p,\xi})=\frac{-\pi \lambda m_D^2}{p(p^2+m_D^2)^2}+\xi \pi \lambda m_D^2\left[ \frac{3  \sin^2\theta_n}{4p(p^2+m_D^2)^2}-\frac{2 m_D^2 (\sin^2\theta_n-\frac{1}{3})}{p(p^2+m_D^2)^3}\right]\,,
\ee
where $\theta_n$ denotes the angle between ${\bf p}$ and the direction of anisotropy ${\bf n}$. Fourier transforming ${\rm Im}\,D^{00}(p_0=0,{\bf p,\xi})$ into coordinate space, the resulting HQ potential can be obtained as
\be\la{vimxa}
{\rm Im}\,V_{\rm pt} (r, \theta, \xi)= \alpha \lambda  [ \phi_2({\hat r}) + \xi (\psi_1(\hat{r},\theta)+\psi_2(\hat{r},\theta))]-\alpha \lambda (1-\xi/6)\, ,
\ee
where  
\ba
\psi_{1}(\hat{r},\theta) &=& -\frac{3}{2} \int_0^\infty \mathrm{d}{z} \frac{z}{(z^2 +1)^2} \bigg(\sin^{2}\theta\frac{\sin(z \hat{r})}{z \hat{r}}+(1-3\cos^{2}\theta)G(\hat{r},z)\bigg)\, ,\nonumber\\
\psi_{2}(\hat{r},\theta) &=& 4\int_0^\infty \mathrm{d}{z} \frac{z}{(z^2 +1)^3} \bigg(\bigg(\frac{2}{3}-\cos^{2}\theta\bigg)\frac{\sin(z \hat{r})}{z \hat{r}}+(1-3\cos^{2}\theta)G(\hat{r},z)\bigg)\, ,\nonumber\\
\ea
with
\ba
G(\hat{r},z)=\frac{\hat{r}z\cos(\hat{r}z)-\sin(\hat{r}z)}{(\hat{r}z)^3}\, .  
\ea
As previously discussed, since no medium effect exists at the origin, one expects
a vanishing imaginary part of the HQ potential in the limit $r\rightarrow 0$. This is actually ensured in eq.~(\ref{vimxa}) by the fact that $\psi_{1}(r \rightarrow 0)=-1/2$ and $\psi_{2}(r \rightarrow 0)=1/3$. On the other hand, neither $\psi_{1}(\hat{r},\theta)$ nor $\psi_{2}(\hat{r},\theta)$ contributes at infinitely large $r$. Therefore, ${\rm Im}\,V_{\rm pt} (r\rightarrow \infty, \xi)=-\alpha \lambda (1-\xi/6)$ depends entirely on the $r$-independent part in the Fourier transform, which can be simply obtained by implementing the replacement $\lambda \rightarrow \lambda (1-\xi/6)$ in the corresponding isotropic result, as given in eq.~(\ref{vinf}).

Following the same strategy that has been adopted in sec.~\ref{real}, an effective expression based on eq.~(\ref{imvdefpt}) for ${\rm Im}\,V_{\rm pt} (r, \theta, \xi)$ is obtained
\be\la{vimpteff}
{\rm Im}\,V_{\rm pt}(r,\theta, \xi) =\alpha \lambda (1-\xi/6) ({\cal I}_2(r {\hat m}_D(\lambda,\xi,\theta))-1)\, .
\ee
Obviously, this effective expression has the desired asymptotic behavior, as discussed above, because ${\cal I}_2(r \rightarrow 0)=1$ and ${\cal I}_2(r \rightarrow \infty)=0$. As we will show later, for the imaginary part of the HQ potential, the anisotropic screening mass is different from that for the real part and therefore is denoted by a different symbol ${\hat m}_D(\lambda,\xi,\theta)$ in eq.~(\ref{vimpteff}).

In the above effective expression, we assume a small modification of the isotropic Debye mass, namely, $({\hat m}_D-m_D) \sim \xi$ and we discard contributions beyond linear order in $\xi$. The matching to the exact result eq.~(\ref{vimxa}) then determines the anisotropic screening mass ${\hat m}_D(\lambda,\xi,\theta)$ as follows
\ba\la{imratio}
\frac{{\hat m}_D}{m_D}&=&1+\xi \frac{{\cal I}_2({\hat r})/6+\psi_{1}(\hat{r},\theta) +\psi_{2}(\hat{r},\theta) }{ {\cal I}_2^\prime({\hat r})\, {\hat r}}+ {\cal O}(\xi^2) \, \nonumber \\
&=&1-\xi \frac{17-9\cos(2\theta)}{120}\, ,\quad{\rm when}\quad {\hat r}\rightarrow 0\, ,
\ea
where $ {\cal I}_2^\prime({\hat r})$ is the derivative of ${\cal I}_2({\hat r})$, which is related to the Meijer's G-function. Clearly, the matching also shows an ${\hat r}$-dependence and, by construction, becomes irrelevant to the explicit form of ${\hat m}_D(\lambda,\xi,\theta)$ in the limiting case when $r \rightarrow 0$ or $r \rightarrow \infty$, where the effective expression eq.~(\ref{vimpteff}) is independent on ${\hat m}_D(\lambda,\xi,\theta)$. 

\begin{figure}
\centering
\includegraphics[width=0.45\textwidth]{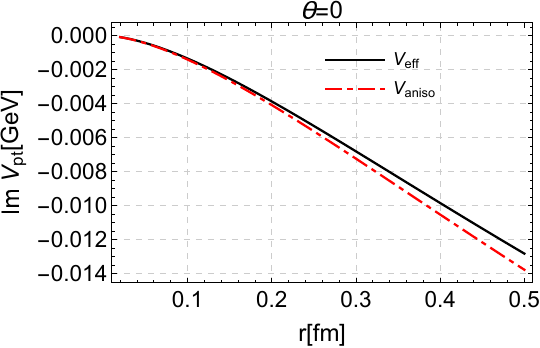}
\includegraphics[width=0.45\textwidth]{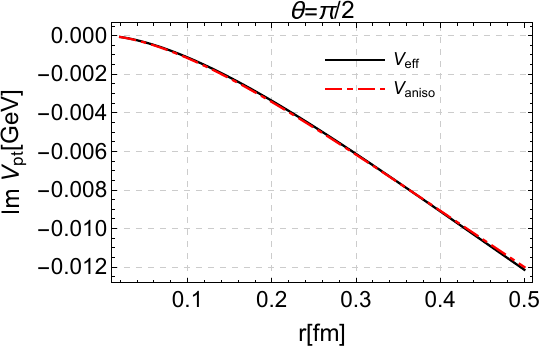}
\vspace*{-0.2cm}
\caption{\label{ibnew}
The imaginary part of the perturbative heavy-quark potential as a function of $r$ evaluated at $\xi=0.7$ and $\theta=0$ (left plot), and $\theta=\pi/2$ (right plot). The solid curves correspond to the exact result while the dash-dotted curves correspond to the effective form eq.~(\ref{vimpteff}) with ${\hat m}_D/m_D = 1-\xi(17-9\cos(2\theta))/120$.  We take $\alpha=0.272$, $\lambda=0.2\,{\rm GeV}$ and $m_D=0.4\,{\rm GeV}$. In this figure, all curves vanish at the origin.}
\end{figure}

\begin{figure}
\centering
\includegraphics[width=0.45\textwidth]{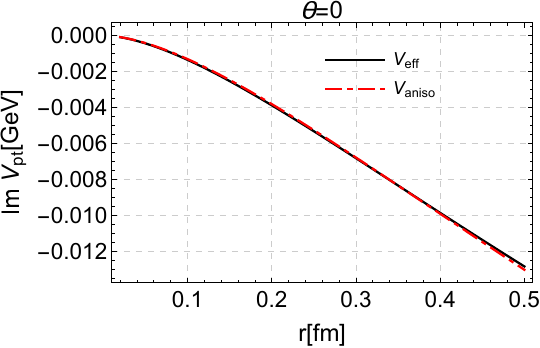}
\includegraphics[width=0.45\textwidth]{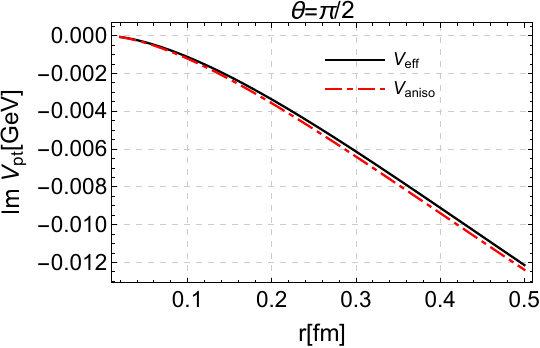}
\vspace*{-0.2cm}
\caption{\label{ib}
The imaginary part of the perturbative heavy-quark potential as a function of $r$ evaluated at $\xi=0.7$ and $\theta=0$ (left plot), and $\theta=\pi/2$ (right plot). The solid curves correspond to the exact result while the dash-dotted curves correspond to the effective form eq.~(\ref{vimpteff}) with ${\hat m}_D/m_D \approx 1-\xi(0.158-0.026 \cos (2\theta))$.  We take $\alpha=0.272$, $\lambda=0.2\,{\rm GeV}$ and $m_D=0.4\,{\rm GeV}$. In this figure, all curves vanish at the origin.}
\end{figure}

When expanding the imaginary part of the perturbative HQ potential in terms of ${\hat r}$, the leading order contributions appear as $\sim {\hat r}^2 \ln {\hat r}$. If the matching between eq.~(\ref{vimxa}) and eq.~(\ref{vimpteff}) is performed at this order, one obtains 
\be\la{aeffim0}
{\tilde m}_D/m_D=1-\xi(17-9\cos(2\theta))/120 \approx 1-\xi(0.142-0.075 \cos (2\theta))\, ,\quad {\rm for}\quad {\hat r}\ll 1\,.
\ee
On the other hand, if we evaluate eq.~(\ref{imratio}) at ${\hat r}=1$, it leads to\footnote{An exact expression can be found, which is quite complicated and, therefore, omitted here.} 
\be\la{aeffim}
{\hat m}_D/m_D \approx 1-\xi(0.158-0.026 \cos (2\theta))\,  ,\quad {\rm for}\quad {\hat r}=1\,.
\ee
Unlike the anisotropic screening mass for the real part of the potential, the determination of ${\hat m}_D(\lambda,\xi,\theta)$ exhibits a strong dependence on ${\hat r}$ since the coefficient of $\cos (2\theta)$ dramatically decreases when increasing the value of ${\hat r}$. As shown in figs.~\ref{ibnew} and \ref{ib}, when ${\hat m}_D(\lambda,\xi,\theta)$ is given by eq.~(\ref{aeffim0}), a less satisfactory reproduction of the exact result is found for $\theta=0$. In order to determine the anisotropic screening masses uniformly for both the real and imaginary part, it turns out that an optimal way is to match the effective expression to the exact perturbative result at ${\hat r}=1$.

Given the above discussions, generalizing the isotropic ${\rm Im}\,V_{\rm npt}(\lambda, r)$ in eq.~(\ref{imvdefnpt}) to the anisotropic QGP is straightforward. Following the same method used to derive eq.~(\ref{rV1}), we arrive at
\ba
\label{iV1}
\mathrm{Im}\,V (r, \theta, \xi) &=& \alpha \lambda (1-\xi/6) ({\cal I}_2(r {\hat m}_D(\lambda,\xi,\theta))-1)+ 4\sigma \lambda \frac{{\cal I}_4(r {\hat m}_D(\lambda,\xi,\theta))-1}{m_D^2 (1-\xi/6)}  \nonumber \\
&=& \alpha \lambda(1-\xi/6) \phi_2(r {\hat m}_D)- \alpha\lambda(1-\xi/6)-\frac{8\sigma \lambda}{m_{D}^2(1-\xi/6)}\phi_3(r {\hat m}_D) \nonumber\\
&+&\frac{24\sigma  \lambda}{m_{D}^2(1-\xi/6)}\phi_4(r {\hat m}_D) -\frac{4\sigma \lambda}{m_{D}^2(1-\xi/6)}\, .
\ea
In the above equation, to generalize ${\rm Im}\,V_{\rm npt}(\lambda, r\rightarrow \infty)=- 4\sigma \lambda/m_D^2$ to nonzero anisotropy, $m_D$ is assumed to be proportional to $\lambda$. This is obviously true in hard-loop perturbation theory where $m_D=\sqrt{N_c/3+N_f/6}\, g \lambda$ \cite{Romatschke:2003ms}. However, in the HQ potential model, $m_D$ should be considered as a non-perturbative quantity for consistency. Following ref.~\cite{Dumitru:2009ni}, we introduce a constant $A\approx 1.4$ to account for all possible non-perturbative effects, namely, $m_D=A \sqrt{N_c/3+N_f/6}\, g \lambda$.\footnote{We take $N_c=3$, $N_f=2$ and $g=1.72$ when numerically solving the Schr{\"o}dinger equation in Secs.~\ref{cb} and \ref{realtime}.} Therefore, using the replacement $\lambda \rightarrow \lambda (1-\xi/6)$ when $r\rightarrow \infty$, we obtain
\be
{\rm Im}\,V(r\rightarrow \infty, \xi)=- \alpha\lambda(1-\xi/6) -\frac{4\sigma \lambda}{m_{D}^2(1-\xi/6)}\, .
\ee

Finally, it should be noted that there are three distinguishable screening scales existing in the anisotropic HQ potential model as given by eqs.~(\ref{rV1}) and (\ref{iV1}). The $\theta$-independent one $m_D(1-\xi/6)$ corresponds to the screening scale at infinitely large $r$. According to eq.~(\ref{vinf}), the asymptotic behavior as $r\rightarrow \infty$ is determined by the $r$-independent part in the Fourier transform, therefore, there is no way to introduce any angular dependence even if an anisotropic gluon propagator $D^{00}(p_0=0,{\bf p,\xi})$ is considered. In addition, the angle-independent screening mass is equal for both the real and imaginary part of the HQ potential. 

The other two anisotropic screening masses depend on the angle $\theta$ and are denoted by ${\tilde m}_D(\lambda,\xi,\theta)$ and ${\hat m}_D(\lambda,\xi,\theta)$ for the real and imaginary part, respectively. An obvious difference between them lies in the coefficients of the $\cos (2\theta)$ term which indicates a less accented angular dependence in the imaginary part. 
Although the corresponding effective expressions with the two $\theta$-dependent screening masses as given in eqs.~(\ref{aeffre}) and (\ref{aeffim}) can quantitatively reproduce the exact results of the perturbative HQ potential, one may still envisage the possibility of having only one $\theta$-dependent screening scale for both $\mathrm{Re}\,V (r, \theta, \xi)$ and $\mathrm{Im}\,V (r, \theta, \xi)$ which would certainly simplify the potential model. Naively, such a realization might be expected because the determination of the two screening masses depends on ${\hat r}$ and there is no fundamental reason to choose any specific ${\hat r}$ as long as its value doesn't exceed $1$. However, as already mentioned before, although an increase of the coefficient of $\cos (2\theta)$ in ${\hat m}_D(\lambda,\xi,\theta)$ can be achieved at smaller ${\hat r}$, even the upper limit $9/120$ cannot reach the value that appears in ${\tilde m}_D(\lambda,\xi,\theta)$, which is larger than $\sim 0.1$. 

Based on the above discussion, it turns out to be necessary to introduce different angle-dependent screening masses for the real and imaginary part of the HQ potential. It is interesting to note that the angle-independent screening mass $m_D(1-\xi/6)$ can be obtained from the angle-dependent ones after performing an average over the solid angle, namely, the following relation holds. 
\be\la{avmd}
\int \frac{d \Omega}{4\pi}  {\tilde m}_D(\lambda,\xi,\theta) = \int \frac{d \Omega}{4\pi}  {\hat m}_D(\lambda,\xi,\theta)=m_D(1-\xi/6)\, .
\ee
The above result is independent on the value of ${\hat r}$ at which ${\tilde m}_D(\lambda,\xi,\theta)$ and ${\hat m}_D(\lambda,\xi,\theta)$ are determined. In fact, the ratio between the anisotropic screening mass and $m_D$ can be uniformly expressed as $1-\xi[1/6+g({\hat r})(1+3\cos(2\theta))]$
where $g({\hat r})$ contains all the ${\hat r}$-dependence as shown in eqs.~(\ref{reratio}) and (\ref{imratio}).\footnote{For Eq.~(\ref{imratio}), $ g(\hat{r})= -\frac{1}{4 {\cal I}_2^\prime({\hat r})\, {\hat r}} \int_0^\infty \mathrm{d}{z} \frac{z^3-5z/3}{(z^2 +1)^3} \big(\frac{\sin(z \hat{r})}{z \hat{r}}+ 3 G(\hat{r},z)\big)$.} Since $\int d \Omega (1+3\cos(2\theta))$ vanishes, the explicit form of $g({\hat r})$ has no influence on the integral in eq.~(\ref{avmd}).

\section{Extremely heavy quarkonia in the anisotropic QGP}
\la{exhq}

To obtain some insight into the anisotropic potential model proposed above, we consider the binding energy for a quarkonium state with extremely large quark mass $M$. As $M\rightarrow \infty$, it is essentially a Coulombic state and the medium effect can be taken as a perturbation. For small ${\hat r}$, eq.~(\ref{rV1}) can be expanded as
\be\la{rex}
{\rm Re}\,V (r, \theta, \xi) = - \frac{\alpha}{r}+\left[-\frac{\alpha}{2}m_D(1-\xi/6)+\frac{\sigma}{m_D(1-\xi/6)}\right] r {\tilde m}_D(\lambda,\xi,\theta)+\cdots\, .
\ee

For the ground state ($1S$), at leading order in the above Taylor series, the eigenenergy of the bound state can be written as $E =-\alpha^2 M/4$. After taking into account eq.~(\ref{rVinf}), the binding energy reads
\be\la{gs}
|E_{\rm bind}|(1S)=\frac{\alpha^2 M}{4}+\frac{2\sigma}{m_D}-\alpha m_D +\frac{\xi}{6}\Big(\alpha m_D+\frac{2\sigma}{m_D}\Big)\, ,
\ee
where contributions beyond linear order in $\xi$ have been neglected. The above result clearly indicates that at finite temperature, the binding of the bound state is weakened, compared to the vacuum case which is a consequence of a decreased ${\rm Re}\,V (r\rightarrow \infty)$ in-medium. Notice that the vacuum potential at infinity should be considered as a large constant due to string breaking. On the other hand, after taking into account the anisotropy effect, an increasing binding is found. We note that the result in eq.~(\ref{gs}) does not depend on ${\tilde m}_D(\lambda,\xi,\theta)$ because it appears at higher order in the small ${\hat r}$ expansion as shown in eq.~(\ref{rex}).

In the anisotropic QGP, the energy splitting of quarkonium states with non-zero angular momentum can be explored by using the perturbation theory in quantum mechanics to compute the leading order correction to the vacuum energy. For this purpose, we can focus on the $1P$ states. The one with angular momentum $L_z=0$ is denoted as $1P_0$, while $1P_{\pm 1}$ refers to the state with $L_z=\pm 1$. Given the anisotropic potential model, the binding energy splitting between $1P_0$ and $1P_{\pm 1}$ is
caused entirely by the difference in their eigenenergies because the potential at infinitely large distance is the same for both states. Therefore, it can be shown that perturbative corrections to the eigenenergy at order $\sim {\hat r}$ or $\sim m_D/(\alpha M)$ lead to a binding energy splitting as the following,
\be\la{sp}
\Delta E_{\rm bind}= |E_{\rm bind}|(1P_0)-|E_{\rm bind}|(1P_{\pm 1})=4 \xi c \frac{\alpha m_D^2-2\sigma}{\alpha M} \, .
\ee
In the above equation, the number $c$ is defined by ${\tilde m}_D/m_D=1+\xi (c \cos (2\theta)+d)$ with $2c-6d=1$ according to the discussion below eq.~(\ref{avmd}). At phenomenologically relevant temperatures, $\alpha m_D^2 < 2\sigma$ can be satisfied. \footnote{For $N_f=2$, the highest temperature that meets this requirement is about $450\, {\rm MeV}$. In general, it becomes irrelevant for quarkonium bound states when considering even higher temperatures.} Thus, the $1P_{\pm 1}$ state is bound more tightly than the $1P_0$ state because $c$ is always positive in our potential model. It is worth noting that without the string contribution, a different conclusion can be drawn, namely, the (absolute value of the) binding energy of $1P_0$ state is larger than that of $1P_{\pm 1}$ state.
Clearly, the anisotropy correction from the string contribution influences the binding of the two $1P$ states in the opposite way as compared to that from the perturbative Debye screened contribution.
However, one has to be very careful to make such a statement because it could be model-dependent. A more detailed discussion on the model dependence can be found in app.~\ref{modde}.

Similarly, for small ${\hat r}$ the imaginary part of the HQ potential model as given in eq.~(\ref{iV1}) can be expanded as 
\be\la{imx}
{\rm Im}\,V (r, \theta, \xi) = \frac{1}{3}\alpha \lambda {\hat r}^2 \ln {\hat r}+ \xi\frac{2}{3} \alpha \lambda {\hat r}^2 \ln {\hat r}\Big(c^\prime \cos (2\theta)+d^\prime-\frac{1}{12}\Big)+\cdots\, 
,\ee
where we assume ${\hat r}^2\ll\xi\ll 1$ and neglect contributions beyond $\sim \xi {\hat r}^2 \ln {\hat r}$. The numbers $c^\prime$ and $d^\prime$ are defined by ${\hat m}_D/m_D=1+\xi (c^\prime \cos (2\theta)+d^\prime)$ and satisfy $2c^\prime-6d^\prime=1$.

Treating the imaginary part as a perturbation of the vacuum Coulomb potential, we can also estimate a decay width for the heavy bound state.\footnote{Since the imaginary part of the HQ potential is negative, we use the absolute values of $\mathrm{Im}\,V (r, \theta, \xi)$  in the Schr{\"o}dinger equation.} The result for the ground state is given by
\ba\la{gw}
\Gamma(1S) &=& \frac{4 \lambda m_D^2}{\alpha M^2} \Big[1-\frac{\xi}{6}(1+4 c^\prime-12 d^\prime)\Big] \ln \frac{\alpha M}{2 m_D}\, \nonumber \\
&=& \frac{4 \lambda m_D^2}{\alpha M^2} \Big(1-\frac{\xi}{2}\Big) \ln \frac{\alpha M}{2 m_D}\, ,
\ea
where the string contribution in the imaginary part of the potential appears at higher order in the small ${\hat r}$ expansion and, therefore, does not show up in the leading order decay width. The above equation suggests that a reduced decay width can be expected when the medium is anisotropic. Unlike the binding energy in eq.~(\ref{gs}), the anisotropic screening mass ${\hat m}_D(\lambda,\xi,\theta)$ which shows up at leading order in the small ${\hat r}$ expansion has been involved in the determination of the perturbative decay width. However, due to the condition $2c^\prime-6d^\prime=1$, the above result doesn't depend on the details of ${\hat m}_D(\lambda,\xi,\theta)$. In other words, no matter at which values of ${\hat r}$ we determine ${\hat m}_D(\lambda,\xi,\theta)$, eq.~(\ref{gw}) is unchanged and also identical to the result obtained by using the exact perturbative HQ potential in eq.~(\ref{vimxa}). 

Furthermore, the splitting of the decay width between the $1P_0$ and $1P_{\pm 1}$ states reads
\be\la{spw}
\Delta \Gamma= \Gamma(1P_0)-\Gamma(1P_{\pm 1})=16 \xi c^\prime \frac{4 \lambda m_D^2}{\alpha M^2}  \ln \frac{\alpha M}{2 m_D}\, .
\ee
Because $c^\prime$ is positive,\footnote{The positivity of $c^\prime$ can be guaranteed in the perturbative region where ${\hat r}\le 1$.} the $1P_0$ state has a larger decay width than the $1P_{\pm 1}$ state. Together with eq.~(\ref{sp}), we find that the bound state with $L_z=\pm 1$ has a higher dissociation temperature as compared to that with $L_z=0$. Obviously, the corresponding results on the energy splitting of the p-wave state depend on the details of the anisotropic screening masses ${\tilde m}_D(\lambda,\xi,\theta)$ and ${\hat m}_D(\lambda,\xi,\theta)$ because of the $c$- or $c^\prime$-dependence. Interestingly, we find that when the two screening masses are given by eqs.~(\ref{aeffre0}) and (\ref{aeffim0}), eqs.~(\ref{sp}) and (\ref{spw}) with vanishing string tension become identical to those obtained by using the exact perturbative HQ potential. However, our qualitative conclusions do not change if the anisotropic screening masses are determined at ${\hat r}=1$. In fact, this is easy to understand. Being essentially a Coulombic state, the properties of the extremely heavy quarkonium is only sensitive to the short distance behavior of the potential which is more accurately reproduced by the effective expression when the matching happens at smaller ${\hat r}$. For the same reason, a more reliable result can be expected for quarkonium states with a relatively large size when eqs.~(\ref{aeffre}) and (\ref{aeffim}) are used for ${\tilde m}_D(\lambda,\xi,\theta)$ and ${\hat m}_D(\lambda,\xi,\theta)$, respectively.

\section{Static in-medium properties of charmonia and bottomonia based on the anisotropic HQ potential model}
\la{cb}

To understand the properties of charmonia and bottomonia in an anisotropic medium, one needs to numerically solve a 3D Schr{\"o}dinger equation with the anisotropic potential model as given in eqs.~(\ref{rV1}) and (\ref{iV1}). To do so, a previously developed code called quantumFDTD will be adopted as the equation solver~\cite{Strickland:2009ft,Delgado:2020ozh}. The obtained eigen/binding energies and decay widths of low-lying quarkonium bound states can be found in app.~\ref{appa} where we used the anisotropic screening masses as determined by eqs.~(\ref{aeffre}) and (\ref{aeffim}) and the anisotropy parameter is set to be $1$. In particular, we consider the ground states $J/\Psi$ and $\Upsilon(1S)$ and also the low-lying p-wave states of bottomonia denoted by $\chi_{b 0}(1P)$ and $\chi_{b \pm1}(1P)$ for $L_z = 0$ and $L_z = \pm1$, respectively.

Compared to the isotropic case, our numerical results show that the magnitudes of the binding energies increase for all the bound states under consideration, which, together with the decreased decay widths, leads to higher dissociation temperatures in the anisotropic QGP. In addition, the removal of the degeneracy of p-wave states gives rise to a larger binding energy in magnitude as well as a smaller decay width for the $\chi_{b \pm1}(1P)$, which therefore is more bound than the $\chi_{b 0}(1P)$. This is actually consistent with our previous analysis of extremely heavy bound states.  
Explicitly, the dissociation temperatures determined by requiring the absolute values of the binding energy equaling twice the decay width are given in Table~\ref{dist}. We point out that although the difference in the dissociation temperatures between the two polarized p-wave states is subtle at $\xi=1$, it could increase for larger $\xi$. However, such an increase is not monotonous with the anisotropy since the two states become degenerate as $\xi \rightarrow \infty$ where the medium density vanishes and the vacuum potential should be used.
 
\begin{table}[htpb]
\begin{center}
\setlength{\tabcolsep}{5mm}{
\begin{tabular}{   |c|  c|  c| }
\hline
$         $ & $\xi=0$ & $\xi=1$     \\ \hline
$  \Upsilon(1S)$ & $338$  & $ 405             $     \\  \hline
$\chi_{b \pm1}(1P)$  & $217$  & $263         $     \\  \hline
$\chi_{b 0}(1P)$  & $217$  & $255       $     \\   \hline	
$J/\Psi$  & $221$  & $265         $     \\
\hline	
\end{tabular}
}
\end{center}
\caption{The dissociation temperatures given in units of {\rm {MeV}} for different quarkonium states at $\xi=0$ and $\xi=1$.
}
\label{dist}
\end{table}

As mentioned before, the angular dependence in the potential model requires solving a 3D Schr{\"o}dinger equation which makes the numerical determination of the binding energies and decay widths of the quarkonium states rather time consuming and much more complicated, compared to the case where a spherically symmetric HQ potential can be used. This is indeed the main obstacle to developing phenomenological applications which include momentum-anisotropy effects. As proposed in our previous work, one possible solution to this difficulty is to employ an angle-averaged effective screening mass ${\cal{M}}_{l m}(\lambda,\xi)$ which by definition reads~\cite{Dong:2021gnb}
\ba\la{effm0}
{\cal{M}}_{l m}(\lambda,\xi)&=&\langle {\rm{Y}}_{l m}(\theta,\phi)| \tilde{m}_D(\lambda,\xi,\theta) | {\rm{Y}}_{l m}(\theta,\phi)\rangle\, ,\nonumber \\
&=&\int_{-1}^{1} d \cos \theta \int_{0}^{2\pi} d \phi {\rm{Y}}_{l m}(\theta,\phi)  \tilde{m}_D(\lambda,\xi,\theta) {\rm{Y}}^*_{l m}(\theta,\phi)\, ,
\ea
and where ${\rm{Y}}_{l m}(\theta,\phi)$ refers to the spherical harmonics with azimuthal quantum number $l$ and magnetic quantum number $m$. The idea is to replace the anisotropic screening mass with ${\cal{M}}_{l m}(\lambda,\xi)$ which recovers the spherical symmetry in the potential model, thus significantly simplifying the numerics. As a result, physical properties of quarkonia in an anisotropic plasma can be obtained by analyzing the bound states in an ``isotropic" medium where the screening scales only depend on $\lambda$ and $\xi$. Notice that although no angular dependence appears in the effective screening mass, the quantum numbers $l$ and $m$ still need to be specified in practical applications. With the concept of the ``most similar state"  introduced in ref.~\cite{Dong:2021gnb}, ${\cal{M}}_{l m}(\lambda,\xi)$ is not universal for all the quarkonium states. In particular, for the s-wave states, the anisotropic screening mass should be replaced by ${\cal{M}}_{00}(\lambda,\xi)$, while for the p-wave states, ${\cal{M}}_{10}(\lambda,\xi)$ and ${\cal{M}}_{11}(\lambda,\xi)$ are used for $P_0$ and $P_{\pm 1}$, respectively.

In ref.~\cite{Dong:2021gnb}, a comprehensive analysis of the validity of using such an effective screening mass is provided where only the real part of the HQ potential is discussed. In this work, we apply the same idea to the imaginary part and the corresponding ${\cal{M}}^{\rm re}_{l m}(\lambda,\xi)$ (which replaces $\tilde{m}_D(\lambda,\xi,\theta)$ in eq.~(\ref{rV1})) and ${\cal{M}}^{\rm im}_{l m}(\lambda,\xi)$ (which replaces $\hat{m}_D(\lambda,\xi,\theta)$ in eq.~(\ref{iV1})) are given by
\ba
\mathcal{M}_{lm}^{\rm re}(\lambda,\xi)&=&\langle Y_{l,m}(\theta,\phi)|\tilde{m}_{D}(\lambda,\xi,\theta)|Y_{l,m}(\theta,\phi)\rangle\approx m_{D}\Big[1+\frac{\xi}{6}\big(0.4312 (3 K_{lm}-1)-1\big)\Big]\, , \nonumber \\ 
\mathcal{M}_{lm}^{\rm im}(\lambda,\xi)&=&\langle Y_{l,m}(\theta,\phi)|\hat{m}_{D}(\lambda,\xi,\theta)|Y_{l,m}(\theta,\phi)\rangle\approx m_{D}\Big[1+\frac{\xi}{6}\big(0.1048 (3 K_{lm}-1)-1\big)\Big]\, , \nonumber \\ \label{eq:mlm} 
\ea
with
\be
K_{lm}=\frac{2l(l+1)-2m^{2}-1}{4l(l+1)-3}\,.
\ee
Explicitly, the effective screening masses relevant in our studies are the following,
\ba\label{MR}
\mathcal{M}_{lm}^{\rm re} (\lambda,\xi)&=& m_{D}
\left\{
\begin{array}{lll}
(1-0.1667\xi)& \;\;\;lm=00 \\
(1-0.1092\xi)& \;\;\;lm=10 \\
(1-0.1954\xi)& \;\;\;lm=11 \\
\end{array}
\right.\, ,\;\;\;\nonumber\\
\mathcal{M}_{lm}^{\rm im} (\lambda,\xi)&=& m_{D}
\left\{
\begin{array}{lll}
(1-0.1667\xi)& \;\;\;lm=00 \\
(1-0.1527\xi)& \;\;\;lm=10 \\
(1-0.1737\xi)& \;\;\;lm=11 \\
\end{array}\,.
\right.
\ea

With an effectively isotropic potential model built by replacing the anisotropic screening masses with the above angle-averaged ones, we solve a 1D Schr\"odinger equation and reproduce both the eigenenergies and decay widths obtained from a direct numerical solution of the 3D Schr\"odinger equation for the same underlying anisotropic potential. In a temperature region relevant for quarkonium studies, our results in app.~\ref{appa} demonstrate that at $\xi=1$ the differences in the eigenenergies are within $\sim 2$ {\rm MeV} while the decay widths can be reproduced to within fractions of an {\rm MeV}. Both results amount to an error of less than $\sim 0.5\%$. Thus, besides the results given in ref.~\cite{Dong:2021gnb}, with the inclusion of an imaginary part in the potential, our work further confirms that the 1D effective potential model could provide an efficient method for including momentum-anisotropy effects.

\begin{figure}
\centering
\includegraphics[width=0.45\textwidth]{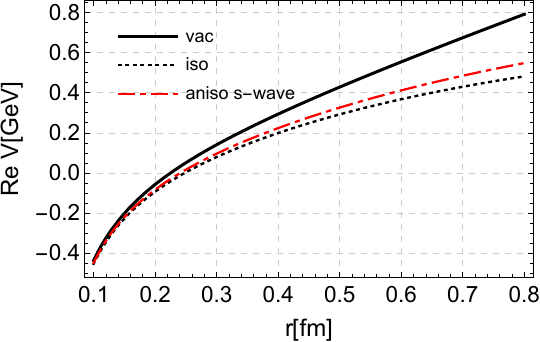}
\includegraphics[width=0.45\textwidth]{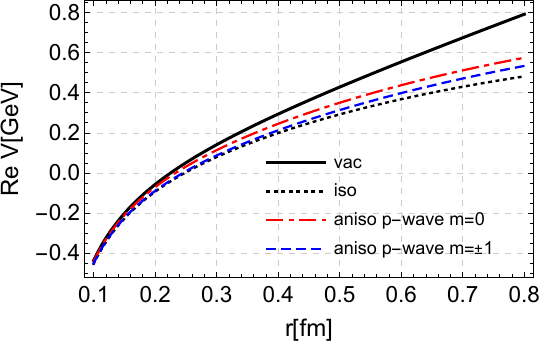}
\vspace*{-0.2cm}
\caption{\label{re}
The real part of the 1D effective potential model as a function of $r$ evaluated at $\xi=1$. The solid curve corresponds to the Cornell potential and the dotted curve is the isotropic potential model as introduced in sec.~\ref{isomod}. We take $\alpha=0.272$,  $\sigma=0.215\,{\rm GeV}^2$ and $m_D=0.5\, {\rm GeV}$. Left: the dash-dotted curve corresponds to the 1D effective potential model with $\mathcal{M}_{00}^{\rm re}$. Right:  the dash-dotted curve and the dashed curve correspond to the 1D effective potential models with $\mathcal{M}_{10}^{\rm re}$ and $\mathcal{M}_{11}^{\rm re}$, respectively.}
\end{figure}

\begin{figure}
\centering
\includegraphics[width=0.45\textwidth]{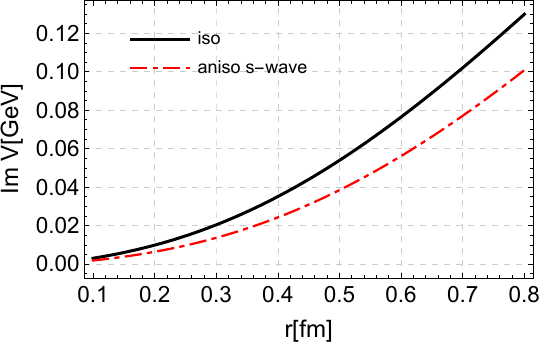}
\includegraphics[width=0.45\textwidth]{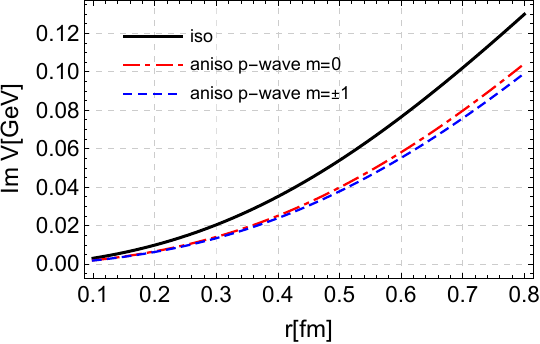}
\vspace*{-0.2cm}
\caption{
The imaginary part of the 1D effective potential model as a function of $r$ evaluated at $\xi=1$. The solid curve corresponds to the isotropic potential model as introduced in sec.~\ref{isomod}. We take $\alpha=0.272$,  $\sigma=0.215\,{\rm GeV}^2$ and $\lambda=0.2 \,{\rm GeV}$ which also determines the Debye mass via the relation $m_D=A  \sqrt{N_c/3+N_f/6}\,g \lambda$. Left: the dash-dotted curve corresponds to the 1D effective potential model with $\mathcal{M}_{00}^{\rm im}$. Right:  the dash-dotted curve and the dashed curve correspond to the 1D effective potential models with $\mathcal{M}_{10}^{\rm im}$ and $\mathcal{M}_{11}^{\rm im}$, respectively.}
\label{im}
\end{figure}

It is also shown that with a fixed anisotropy, the eigenenergy of a quarkonium state increases with decreasing temperature which is in accordance with the fact that the HQ potential is getting closer and closer to the vacuum Cornell potential as we decrease the temperature. For the same reason, one can also expect that at a given temperature, the anisotropic potential overshoots the isotropic one, in other words, the former is closer to the Cornell potential because the eigenenergy increases when an anisotropy is present. In addition, the eigenenergy of $\chi_{b0}(1P)$ is found to be larger than that of $\chi_{b\pm1}(1P)$ which also suggests that the interaction between the quark and anti-quark that make up the $P_0$ state should be described by a potential closer to the vacuum Cornell potential. As a consequence, the $P_0$ state is less bound than the $P_{\pm1}$ state because by construction the potential at infinitely large distance is the same for these two states. These statements can be understood more clearly when we look at the 1D effective potential models with different quantum numbers $l$ and $m$ in the angle-averaged mass $\mathcal{M}_{lm}^{\rm re} (\lambda,\xi)$ which are plotted in fig.~\ref{re}. A similar conclusion also applies for the imaginary part of the potential. As shown in fig.~\ref{im}, it decreases in magnitude when non-zero anisotropy is present and the potential corresponding to $\chi_{b\pm1}(1P)$ is deeper than that for $\chi_{b0}(1P)$.

\section{Real-time solution to the Schr\"odinger equation}
\label{realtime}

In this section we consider the real-time solution of the Schr\"odinger equation.  To model the real-time evolution with the angle-dependent heavy-quark potential \eqref{rV1} and \eqref{iV1} we solve the 3D Schr\"odinger equation on a lattice with $N=128$ points in each direction. For bottomonium states we take the box size to be $L =$ 2.56 fm and use $m_b = 4.7$ GeV.\footnote{We report results for charmonium in app.~\ref{app:charmonium}.}  To evolve the 3D wave function for a given potential we use a split-step pseudo-spectral splitting method~\cite{Fornberg:1978,Taha:1984jz} with temporal step size $\Delta t = 0.001$ fm/c.  In all results presented in this section we evolve the wave function from $\tau = 0$ fm/c to $\tau = 0.25$ fm/c in the vacuum ($T=0$) heavy-quark potential which corresponds to a Cornell potential.  Starting at $\tau = \tau_0 = 0.25$ fm/c, we assume a fixed anisotropy parameter $\xi = 1$ and boost-invariant Bjorken evolution for the hard scale
\be
\lambda(\tau) = \lambda_0 \left( \frac{\tau_0}{\tau} \right)^{1/3} \, .
\ee
We take the initial hard scale to be $\lambda_0 =$ 630 MeV, which corresponds to initial temperatures achieved in central 5.02 TeV Pb-Pb collisions at the LHC~\cite{Alqahtani:2020paa}.

\subsection{Numerical method}

To obtain the split-step update rule for the wave function, one decomposes the Hamiltonian into kinetic and potential contributions, $H = T + V$, and makes use of the Baker-Campbell-Hausdorff theorem to obtain
\begin{equation}
\exp\left[- i H \Delta t\right] \simeq \exp\left[- i V \frac{\Delta t}{2} \right] \exp\left[- i T \Delta t\right]  \exp\left[- i V \frac{\Delta t}{2} \right] + \mathcal{O}((\Delta t)^2) \, .
\end{equation}
Since the kinetic energy only depends on momentum and the potential only depends on position, one can perform the corresponding updates directly in momentum or position space. Based on this understanding, the 3D wave function can be updated a single step $\Delta t$ forward in time as follows
\begin{itemize}
\setlength{\itemindent}{-3mm}
	\item Update the wave function in configuration space using a half-step: $ \psi_1 = \exp\!\left[- i V \frac{\Delta t}{2} \right] \psi_0$.
	\item Fourier transform the wave function: $\tilde\psi_1 = \mathbb{F}_s[\psi_1]$.
	\item Update the wave function in momentum space using: $\tilde\psi_2 =  \exp\!\left[- i T \Delta t\right] \tilde\psi_1$.
	\item Apply an inverse Fourier transform: $\psi_2 = \mathbb{F}_s^{-1}[\tilde\psi_2]$.
	\item Finally, update the wave function in configuration space using a half-step:\\$\psi_3 = \exp\!\left[- i V \frac{\Delta t}{2} \right] \psi_2$.
\end{itemize}
For real-valued potentials, this method is manifestly unitary.
In order to optimize code speed we make use of GPU-acceleration provided by the CUDA parallel computing platform and parallelize the necessary 3D Fast Fourier Transforms using the CUDA cuFFT package~\cite{cuda}.  Use of massive parallelization allows us to simulate large 3D lattices more efficiently than MPI-based implementations.  For the results presented here we used a single Tesla K20m with 2496 cores, which allows for lattice sizes of up to $256^3$ using double precision.

To set the initial condition for the real-time evolution, we consider two possibilities:  (1) initializing with an eigenstate of the vacuum Hamiltonian or (2) initializing with a localized Gaussian wave function.  For the first case we determine the 3D vacuum eigenstates by discretizing the Hamiltonian in the coordinate basis on a 3D lattice that matches the one used for the 3D real-time evolution.  For this purpose we implemented an MPI-based CPU code, which uses the PETSC and SLEPC libraries to determine the eigenvalues and eigenstates \cite{petsc,slepc}. The standard Krylov-Schur method of SLEPC is used with a tolerance of $\Delta=10^{-15}$, allocating twice the number of workspace memory as eigenvectors requested. No preconditioning was found to be necessary. The 3D eigenstates are saved to disk in binary format and then read into the CUDA-based real-time solver {\tt s3d-cuda}.  We note that, to match the discretization of the derivative used in the vacuum eigensolver, in {\tt s3d-cuda} we use a kinetic term $T = 2[1-\cos(pa)]/(ma^2)$ with $a$ being the lattice spacing, which corresponds to using the standard second order accurate discrete second derivative approximation.  In order to monitor the evolution of the wave function, we report the time evolution of the probability of the system being in a particular vacuum eigenstate.  This is determined by computing the ``overlap'' integral
\be
p_i = \left|\int d^3{\bf x} \, \psi_i^*({\bf x}) \psi(t,{\bf x})\right|^2 \, ,
\ee
where $i$ indexes the vacuum eigenstates.

\subsection{Results -- Eigenstate initialization}

\begin{figure}[t]
\centering
\includegraphics[width=0.9\textwidth]{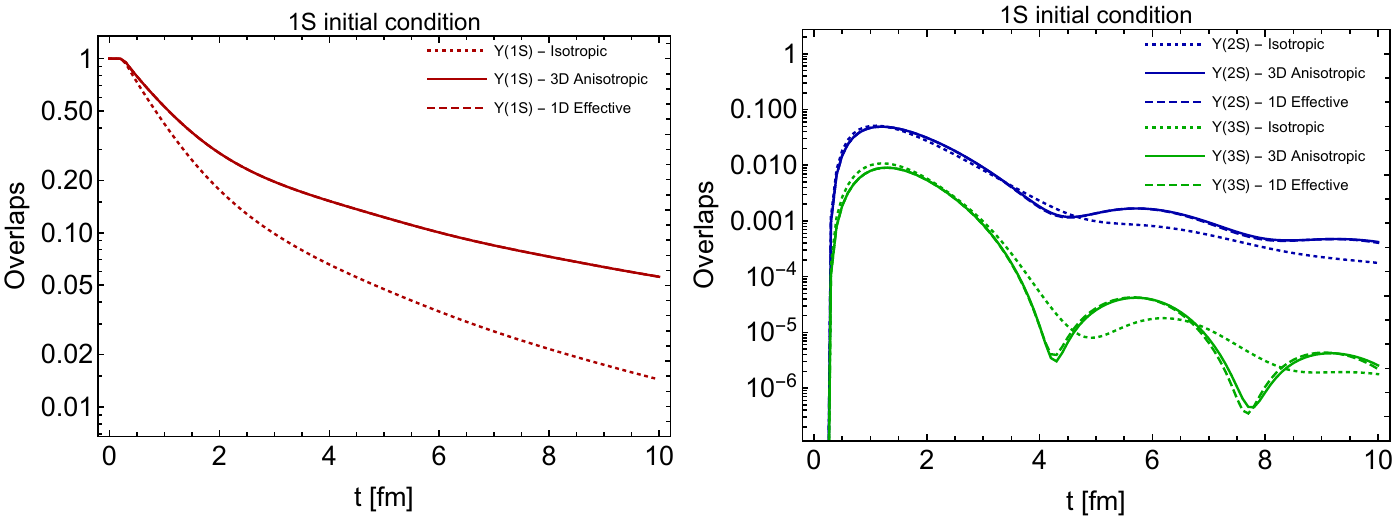}\\[1em]
\includegraphics[width=0.9\textwidth]{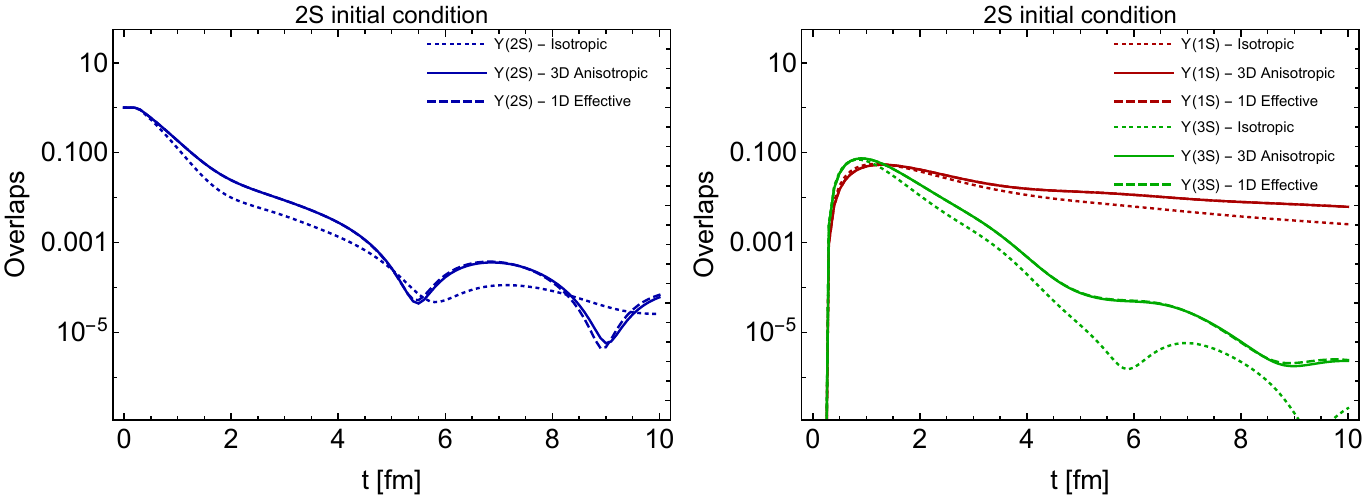}\\[1em]
\includegraphics[width=0.9\textwidth]{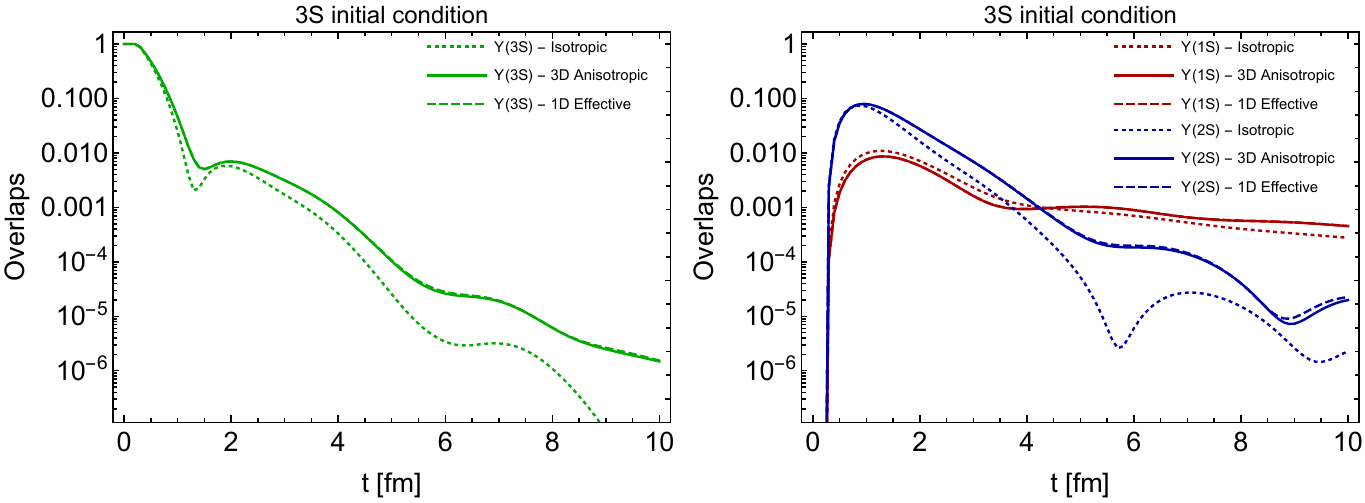}
\caption{$\Upsilon(1S)$, $\Upsilon(2S)$, and $\Upsilon(3S)$ overlaps resulting from real-time solution of the Schr\"odinger equation.  In the top, middle, and bottom rows we initialized the wave function as pure $\Upsilon(1S)$, $\Upsilon(2S)$, and $\Upsilon(3S)$ eigenstates, respectively.  The left panels correspond to diagonal overlaps and the right panels correspond to off-diagonal s-wave overlaps.  In each panel there are three line types:  Dotted lines were obtained with an isotropic 1D thermal potential, $T(\tau)=\lambda(\tau)$ and $\xi=0$; solid lines were obtained with the full angle-dependent 3D potential; and, finally, dashed lines were obtained using the 1D effective potential.}
\label{fig:overlaps-swave}
\end{figure}

\begin{figure}
\centering
\includegraphics[width=1\textwidth]{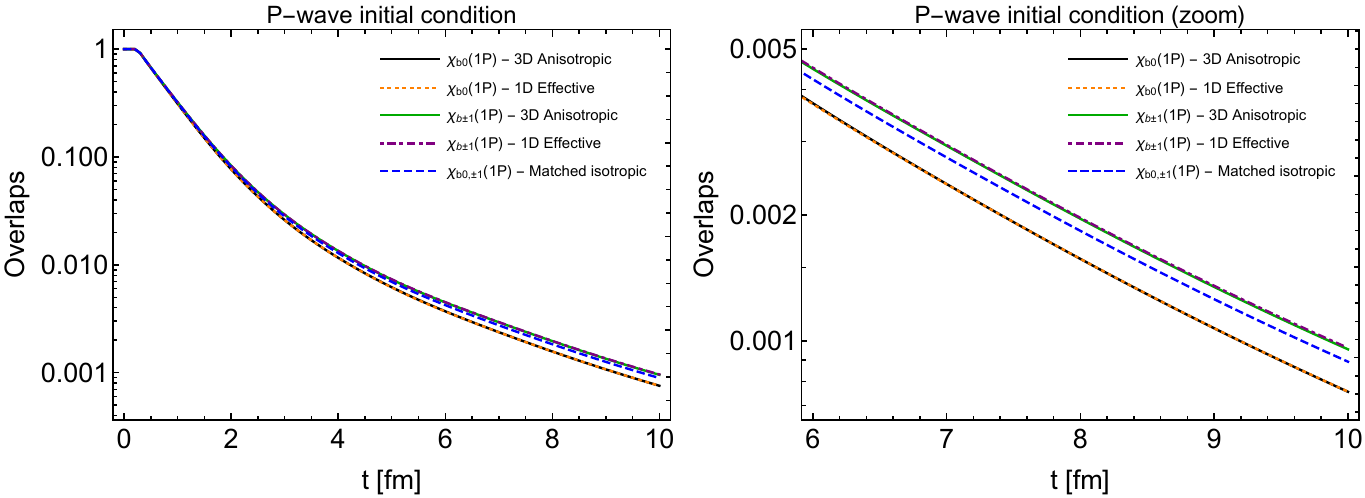}
\vspace*{-0.5cm}
\caption{Time evolution of the bottomonium p-wave overlaps resulting from initialization with different p-wave polarizations corresponding to $l=1$ and $m=0,\pm1$ labeled as $\chi_{b0}(1P)$ and $\chi_{b\pm1}(1P)$, respectively.  The solid black and orange lines correspond to the full 3D evolution with $\chi_{b0}(1P)$ and $\chi_{b\pm1}(1P)$ initial conditions and the orange dotted and purple dot dashed lines correspond to the 1D effective potential evolution with the same initial conditions. The dashed blue line corresponds to the isotropic matching scheme detailed in the text.}
\label{fig:overlaps-pwave}
\end{figure}

In fig.~\ref{fig:overlaps-swave} we present results for the time evolution of the $\Upsilon(1S)$, $\Upsilon(2S)$, and $\Upsilon(3S)$ overlaps for bottomonium states obtained by using pure vacuum eigenstates as the initial condition.  In the top, middle, and bottom rows we initialized the wave function as pure $\Upsilon(1S)$, $\Upsilon(2S)$, and $\Upsilon(3S)$ eigenstates, respectively.  The left panels correspond to the diagonal overlaps and the right panels correspond to the off-diagonal s-wave overlaps.  In each panel there are three line types:  Dotted lines were obtained with an isotropic 1D thermal potential, $T(\tau)=\lambda(\tau)$ and $\xi=0$; solid lines were obtained with the full angle-dependent 3D potential; and, finally, dashed lines were obtained using the 1D effective potential.  In all cases we find that the 1D effective potential provides an excellent approximation to the full 3D anisotropic evolution and that ignoring the momentum anisotropy results in much worse results, particularly for the diagonal overlaps shown in the left panels.  For the diagonal $\Upsilon(1S)$ overlap we find, in particular, that the maximum difference between 3D and 1D effective potential evolution is 0.05\%.  In the case of the $\Upsilon(2S)$ and $\Upsilon(3S)$ diagonal overlaps and $\Upsilon(1S)$, $\Upsilon(2S)$, and $\Upsilon(3S)$ off-diagonal overlaps somewhat larger differences between the 3D and effective 1D evolutions are seen, however, we still find that the 1D effective potential evolution provides an excellent approximation to the full 3D evolution.

Turning to p-wave initial conditions, in fig.~\ref{fig:overlaps-pwave} we present the time evolution of the p-wave overlaps resulting from initialization with different p-wave polarizations corresponding to $l=1$ and $m=0,\pm1$ labeled as $\chi_{b0}(1P)$ and $\chi_{b\pm1}(1P)$, respectively, in the figure.  The solid black and orange lines correspond to the full 3D evolution with $\chi_{b0}(1P)$ and $\chi_{b\pm1}(1P)$ initial conditions.  The orange dotted and purple dot dashed lines correspond to the 1D effective evolution with the same initial conditions.  In this figure we also include the result obtained when ignoring the $l$ and $m$ dependence of the effective masses ${\cal M}^{\rm re}_{lm}$ and ${\cal M}^{\rm im}_{lm}$ as the ``matched isotropic'' case.  In this case we use the effective masses corresponding to $l=0$ and $m=0$, in which case $K_{lm} = 1/3$ and 
${\cal M}^{\rm re}_{lm}={\cal M}^{\rm im}_{lm}=m_D ( 1 - \xi/6 )$. At leading order in $\xi$ this mass prescription is equivalent to Landau matching the effective isotropic and underlying anisotropic energy density, $\epsilon(T_{\rm eff}) = \epsilon(\lambda,\xi)$
\cite{Margotta:2011ta} which results in $T_{\rm eff} = \lambda (1 - \xi/6)$.

As can be seen from fig.~\ref{fig:overlaps-pwave}, the 1D effective evolution of the polarized p-wave overlaps agrees well with the result obtained using the full 3D anisotropic potential.  This includes the splitting between the $m=0$ and $|m|=1$ initial conditions due to the presence of momentum anisotropy in the system.  We note that the matched isotropic result cannot describe this splitting since it ignores the $l$ and $m$ dependence of the effective masses.  Taking into account the $l$ and $m$ dependence using the 1D effective potential therefore represents a substantial improvement in our description of the time evolution of higher angular momentum states.

\subsection{Results -- Gaussian initialization}

In practice one does not expect the initial condition for quarkonium to be a pure eigenstate.  In the large quark mass limit one expects to have local production of quarkonium states which maps to a delta function in coordinate space \cite{Laine:2007gj,Brambilla:2020qwo}.  Finite mass effects result in a broadening of this delta function with a width on the order of $1/M$.  In order to model this we use a Gaussian-smeared delta function of the form
\begin{figure}
\centering
\includegraphics[width=0.8\textwidth]{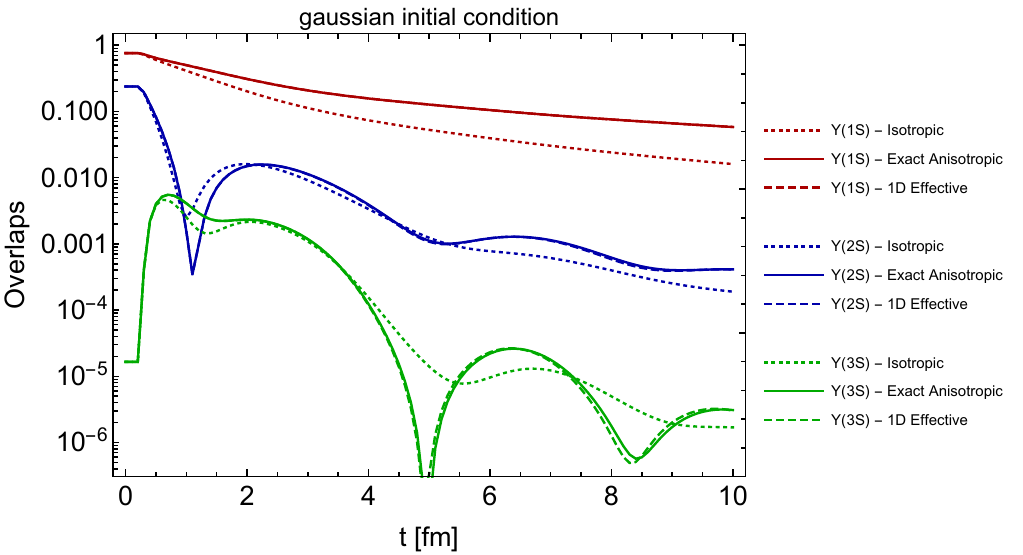}
\caption{Time evolution of the $\Upsilon(1S)$, $\Upsilon(2S)$, and $\Upsilon(3S)$ overlaps.  The dotted, solid, and dashed lines correspond to using the 1D isotropic, 3D anisotropic, and 1D effective potentials, respectively.  The red, blue, and green colors correspond to the $\Upsilon(1S)$, $\Upsilon(2S)$, and $\Upsilon(3S)$ overlaps, respectively.}
\label{fig:overlaps-gaussian}
\end{figure}
\be
\psi(\tau_0,{\bf x}) = N \exp(-{\bf x}^2/2 \varsigma^2) \, ,
\label{eq:gaussianform}
\ee
where $\varsigma = 0.354$ fm and $N$ is the normalization which is determined numerically.  This initial condition is a quantum superposition of both bound ($E_\text{bind}<0$) and unbound ($E_\text{bind}>0$) states.  Here we will report on the time evolution of the bound state overlaps.  In fig.~\ref{fig:overlaps-gaussian} we present the time evolution of the $\Upsilon(1S)$, $\Upsilon(2S)$, and $\Upsilon(3S)$ overlaps.  The dotted, solid, and dashed lines correspond to using the 1D isotropic, 3D anisotropic, and 1D effective potentials, respectively.  The red, blue, and green colors correspond to the $\Upsilon(1S)$, $\Upsilon(2S)$, and $\Upsilon(3S)$ overlaps, respectively.  As can be seen from this figure the 1D effective model describes the evolution obtained using this initial condition quite well, whereas ignoring the anisotropy fails to provide a good description of the overlaps, in particular for the 1S overlap.  We note that one can also initialize different polarized p-wave Gaussian initial conditions by multiplying eq.~\eqref{eq:gaussianform} by $x$, $y$, and $z$, finding similar results, namely that the 1D effective potential provides an accurate approximation to the full 3D evolution.

\section{Conclusions}
\la{con}

In this work, we generalized a complex HQ potential model proposed in ref.~\cite{Guo:2018vwy} from an equilibrium QGP setting to a non-equilibrium QGP with small momentum-space anisotropy. The non-equilibrium effects were incorporated into the potential model by utilizing an anisotropic screening mass, which depends on the quark pair alignment with respect to the direction of anisotropy. In practice, this generalization consists of replacing the isotropic Debye mass $m_D$ with the anisotropic screening mass, which we consider as a ``minimal" extension to non-zero anisotropy. By assuming that the very same screening scale appearing in the perturbative contributions also shows up in the non-perturbative string part, the anisotropic screening mass can be determined based solely on the information of the Debye-screened HQ potential which is calculable within hard-thermal-loop perturbation theory even in the non-equilibrium case. 

We applied the above idea to both the real and imaginary part of the HQ potential to derive the anisotropic screening masses, which required matching the effective expressions given by eqs.~(\ref{vpteff}) and (\ref{vimpteff}) to the exact results from a first-principles calculation. These effective expressions were chosen such that the desired asymptotic behavior is automatically guaranteed. The resulting angle-dependent anisotropic screening masses, as determined, had an ambiguity due to the ${\hat r}$-dependence and numerical results showed that matching at ${\hat r}=1$ turned out to be an optimal method which can nearly perfectly reproduce the exact perturbative results with the effective forms. Given the ``minimal" extension, this clearly demonstrated the validity of the key ingredients, namely, the anisotropic screening masses, in our model construction. In addition, we found that it was necessary to introduce different anisotropic screening masses ${\tilde m}_D(\lambda,\xi,\theta)$ and ${\hat m}_D(\lambda,\xi,\theta)$ for the real and imaginary part, respectively, because the angular dependence was much less pronounced in ${\hat m}_D(\lambda,\xi,\theta)$ when compared to that obtained for the real part of the potential.

An important advantage of the anisotropic HQ potential model developed in this work is that it is realized by employing an angle-averaged effective screening mass as defined by eq.~(\ref{effm0}). Replacing the anisotropic screening mass with the angle-averaged one, the numerical solution of the Schr{\"o}dinger equation with a 3D anisotropic HQ potential reduces to a much simpler 1D problem. This in turn allows us to overcome one central obstacle in phenomenological applications, where momentum-anisotropy effects need to be considered. Due to the absence of first principle lattice simulations of the anisotropic HQ potential, it needs to be pointed out that the focus should not be on the validity of the potential model based on eq.~(\ref{anisodis}) itself, but on the method for reduction from 3D to 1D. The success of using such a 1D effective screening mass has been demonstrated rather conclusively in both static and dynamic cases. We reproduced the full 3D results for the binding energies and decay widths of low-lying quarkonium bound states to very high accuracy, {\em i.e.}, with an error of less than $\sim 0.5\%$. For the latter case, by assuming a boost-invariant Bjorken evolution of the medium, we studied the time-dependent probability of the system being in a particular vacuum eigenstate. With different initial conditions for the wave function, all the obtained ``overlaps" from the 1D effective potential were in very good agreement with the full 3D evolution. It is worth noting that in both cases, the splitting of the p-wave states with different polarizations, which emerged as a unique feature in the anisotropic QGP, was also well described by our method.

Although our discussions herein were limited to quarkonium physics, the modeling proposed in this work is expected to have application in many other areas which involve a momentum-space anisotropy in the distribution function. In fact, with the inclusion of a string contribution, the full gluon propagator by which we defined the complex HQ potential encodes essential information on the screening and damping of the anisotropic QGP through the effectively 1D mass scales as given in eq.~(\ref{eq:mlm}). From the phenomenological point of view, such a propagator resums the gluon self-energy in a non-perturbative manner according to the Dyson-Schwinger equation. As is well known, both the gluon self-energy and the propagator are key ingredients in computing many physical processes including those mentioned in the introduction. Therefore, the current work can offer an efficient strategy to non-perturbatively investigate new phenomena associated with momentum-space anisotropy, which in turn could further verify the validity of our modeling.

Finally, we emphasize that it is important to take into account the time evolution of the anisotropy parameter $\xi$ which may become large during the expansion of the fireball created in the heavy ion collisions. Therefore, to fully understand the in-medium dynamics of heavy quarkonium, it is necessary to solve a 3D stochastic Schr{\"o}dinger equation beyond the small $\xi$ approximation.  To do so, one needs to construct a complex HQ potential model for arbitrary anisotropies and assess the applicability of the angle-averaged effective screening masses at large $\xi$, especially for the imaginary part. These lines of inquiry provide interesting paths forward for future work.

\section*{Acknowledgements}
The work of Y.G. is supported by the NSFC of China under Project No. 12065004 and 12147211. M.S. and A.I.  were supported by the U.S. Department of Energy, Office of Science, Office of Nuclear Physics Award No.~DE-SC0013470. A.R. gladly acknowledges support from the Research Council of Norway under the FRIPRO Young Research Talent grant 286883 and from UNINETT Sigma2 - the National Infrastructure for High Performance Computing and Data Storage in Norway under project NN9578K-QCDrtX "Real-time dynamics of nuclear matter under extreme conditions". 

\appendix

\section{Model dependence of the splitting of p-wave states in the anisotropic QGP}
\la{modde}

The complex HQ potential model for non-zero anisotropy as proposed in this work consists of two contributions. The validity of the Debye screened contribution can be assessed by a direct comparison with the exact perturbative computations. However, due to a lack of information from first principle calculations, there exist ambiguities with regard to modeling the string contribution. The ``minimal" extension based on an isotropic potential model strongly depends on one basic assumption, as already mentioned in sec.~\ref{apot}. Even so, there exist alternative ways to model the anisotropic HQ potential, which may lead to a different conclusion on the splitting of the p-wave states.

Considering the $r$-independent contribution in the HQ potential, the generalization to nonzero anisotropy is merely a replacement $m_D\rightarrow m_D(1-\xi/6)$. As for the $r$-dependent contribution $\alpha m_D\,{\cal I}_1({\hat r})$ in eq.~(\ref{revdefpt}), a new effective expression given by $-\alpha e^{-r {\tilde m}_D(\lambda,\xi,\theta)}/r$ may also be considered. Here, all the $m_D$'s are treated equivalently and a $\theta$-dependence is introduced in the screening mass when $\xi \neq 0$. In contrast, only those $m_D$'s in ${\cal I}_1({\hat r})$ have been replaced by ${\tilde m}_D(\lambda,\xi,\theta)$ in the effective expression eq.~(\ref{vpteff}). At first glance, there seems to exist a better justification, if the perturbative potential remains of Debye screened form, even in an anisotropic medium so that the anisotropic screening mass ${\tilde m}_D(\lambda,\xi,\theta)$ has a clear physical meaning. However, further investigation shows that when the $r$-independent contribution is included, the resulting ${\rm Re}\,V_{\rm pt}(r,\theta, \xi)$ does not reproduce the vacuum Coulomb potential when $r\rightarrow 0$, because an extra $r$-independent term $\alpha ({\tilde m}_D-m_D(1-\xi/6))$ emerges in this limit. 

Despite the above mentioned defects, we can also determine the anisotropic screening mass ${\tilde m}_D(\lambda,\xi,\theta)$ in the same way as before which gives
\ba\la{reratio2}
\frac{{\tilde m}_D}{m_D}
&=&1+\xi \Big(-\frac{1}{6}+ \frac{6(1-e^{ {\hat r}})+ {\hat r}(6+3 {\hat r}+ {\hat r}^2)}{12  {\hat r}^3}(1+3\cos(2\theta)) \Big)\, \nonumber \\
&=&1+\xi \Big(-\frac{1}{6}-\frac{1}{48}(1+3\cos(2\theta)){\hat r} +\cdots \Big)\, .
\ea
The above equation becomes ${\tilde m}_D/m_D=1-\xi/6$ at ${\hat r}= 0$ which indicates the reproduction of the vacuum Coulomb potential as $r \rightarrow 0$ due to the vanishing of the above mentioned term. However, without a $\theta$-dependence in the anisotropic screening mass, the resulting effective expression fails to describe the central and unique feature of the anisotropic HQ potential, namely, $V (r,\theta, \xi)$ is dependent on the alignment of the quark pair. In addition, when evaluating the expanded result at ${\hat r}=1$, we find ${\tilde m}_D/m_D=1-\xi(3+\cos (2\theta))/16$ which is identical to that originally obtained in ref.~\cite{Dumitru:2009ni}. As compared to eq.~(\ref{aeffre}), the most notable difference is the sign change of the coefficient of $\cos(2\theta)$. Even without further analysis, one can conclude that the new effective form of ${\rm Re}\,V_{\rm pt}(r,\theta, \xi)$ results in a more bound $P_0$ state because as compared to the case where $\theta=\pi/2$, a reduced screening mass is found when the quark pair is aligned along the direction of the anisotropy.\footnote{Although the same conclusion holds, one cannot apply the discussion equally to the effective form eq.~(\ref{aeffre}) because it doesn't have a standard Debye screened form.}

According to the above discussion, the ``minimal" extension leads to a new form of the anisotropic HQ potential model as the following,
\ba
\label{rV1old}
\mathrm{Re}\,V (r, \theta, \xi) &=&- \alpha \frac{e^{-r {\tilde m}_D}}{r}-\alpha m_D(1-\xi/6)-\frac{2 \sigma}{{\tilde m}_D}e^{-r {\tilde m}_D}(1+\frac{r {\tilde m}_D}{2})+\frac{2 \sigma}{m_D(1-\xi/6)}\nonumber \\
&=& - \frac{\alpha}{r}+ \alpha {\tilde m}_D-\alpha m_D(1-\xi/6)-\frac{2\sigma }{ {\tilde m}_D}+\frac{2\sigma}{m_D(1-\xi/6)}+\cdots\, .
\ea
In the second line of the above equation, we expand the result for small ${\hat r}$ which can be used to study the binding energies for a quarkonium state with extremely large quark mass $M$. For the ground state, including contributions at order $\sim {\hat r}^0$, we find that 
\ba\la{gsold}
|E_{\rm bind}|(1S)&=&\frac{\alpha^2 M}{4}+\frac{2\sigma}{m_D}-\alpha m_D +\frac{\xi}{6}\Big(\alpha m_D+\frac{2\sigma}{m_D}\Big)(2 c-6 d)\,\nonumber \\
&=&\frac{\alpha^2 M}{4}+\frac{2\sigma}{m_D}-\alpha m_D +\frac{\xi}{6}\Big(\alpha m_D+\frac{2\sigma}{m_D}\Big)\, ,
\ea
where the numbers $c$ and $d$ are defined by ${\tilde m}_D/m_D=1+\xi (c \cos (2\theta)+d)$. Different from eq.~(\ref{gs}), the calculation of eq.~(\ref{gsold}) depends on ${\tilde m}_D(\lambda, \xi,\theta)$. However, due to the fact that $2 c-6 d= 1$ which can be obtained from Eq.~(\ref{reratio2}), the above result is identical to eq.~(\ref{gs}).

As for the binding energy splitting between the $1P_0$ and $1P_{\pm 1}$ states, it can be shown that
\be\la{spnew}
\Delta E_{\rm bind}= |E_{\rm bind}|(1P_0)-|E_{\rm bind}|(1P_{\pm 1})=-4 \frac{\alpha m_D^2+2\sigma}{5 m_D} \xi c\, .
\ee
As long as ${\tilde m}_D(\lambda, \xi,\theta)$ is not determined at the origin where no $\theta$-dependence exists, the number $c$ is always negative as indicated by eq.~(\ref{reratio2}). As a result, the conclusion that $1P_0$ state is bound more tightly than $1P_{\pm 1}$ can be drawn which is opposite to what has been obtained based on eq.~(\ref{rV1}). With the new potential model as given in eq.~(\ref{rV1old}), both the perturbative and non-perturbation terms make the binding of $1P_0$ stronger than $1P_{\pm 1}$. In addition, different from eq.~(\ref{sp}), the splitting appears at order $\sim {\hat r}^0$ in eq.~(\ref{spnew}), therefore, terms linear in ${\hat r}$ are not listed in the expansion in eq.~(\ref{rV1old}).

Proposed by the same reasoning as the real part, a new form of the imaginary part of the anisotropic HQ potential model can be expressed as
\ba
\label{iV1newmodel}
\mathrm{Im}\,V (r, \theta, \xi) &=& \alpha {\hat \lambda} \phi_2(r {\hat m}_D)- \alpha\lambda(1-\xi/6) -\frac{4\sigma \lambda}{m_{D}^2(1-\xi/6)}\nonumber \\
&-&\frac{8\sigma {\hat \lambda}}{{\hat m}_{D}^2}\big(\phi_3(r {\hat m}_D)-3\phi_4(r {\hat m}_D)\big)\nonumber\\
&=& \frac{1}{3}\alpha \lambda {\hat r}^2 \ln {\hat r}+ \xi \lambda \frac{\alpha m_D^2-4\sigma}{m_D^2}\Big(\frac{1}{6}+ c^\prime \cos (2\theta)+ d^\prime\Big)\nonumber \\
&+& \alpha  \xi \lambda   {\hat r}^2 \ln {\hat r}(c^\prime \cos (2\theta)+d^\prime)+\cdots \, .
\ea
It is easy to show that the above form of ${\rm Im}\,V (r,\theta, \xi)$ does not vanish at the origin which is obviously an undesired behavior. In the above equation, we also expand $\mathrm{Im}\,V (r, \theta, \xi)$ for small ${\hat r}$ and $\xi$ and keep only the first three terms under the hierarchy ${\hat r}^2 \ll \xi \ll {\hat r}^2 \ln {\hat r} \ll 1$. As before, $ {\hat m}_D/m_D=1+\xi (c^\prime \cos (2\theta)+d^\prime)$ with ${\hat m}_D(\lambda,\xi,\theta)=A \sqrt{N_c/3+N_f/6} \,g {\hat \lambda}(\xi,\theta)$. We have checked that matching the perturbative part in eq.~(\ref{iV1newmodel}) to the exact anisotropic potential leads to the same condition $2c^\prime-6 d^\prime=1$ and $c^\prime$ is negative in the perturbative region where ${\hat r}\le 1$.

Considering the decay width for the ground state, a similar calculation leads to the following result
\ba\la{gsw2}
\Gamma(1S)&=& \frac{4 \lambda m_D^2}{\alpha M^2} \ln \frac{\alpha M}{2 m_D}-\xi \frac{\lambda}{6} \Big[\frac{\alpha m_D^2- 4\sigma}{m_D^2}(1-2c^\prime+6 d^\prime)+\frac{12 m_D^2}{\alpha M^2} \ln \frac{\alpha M}{2 m_D}(2c^\prime-6 d^\prime)\Big]\,\nonumber \\
&=& \frac{4 \lambda m_D^2}{\alpha M^2} \Big(1-\frac{\xi}{2}\Big) \ln \frac{\alpha M}{2 m_D}\, .
\ea
As compared to eq.~(\ref{gw}), in both cases the medium correction at zero anisotropy arises at order $\sim {\hat r}^2 \ln {\hat r}$. However, in eq.~(\ref{gw}) the anisotropy effect comes in at order $\sim \xi {\hat r}^2 \ln {\hat r}$ which becomes sub-leading in the above equation as the leading order contribution with non-zero anisotropy shows up at order $\sim \xi {\hat r}^0$. On the other hand, thanks to the condition $2c^\prime-6 d^\prime=1$, the contribution $\sim \xi {\hat r}^0$ vanishes and the above result is actually identical to eq.~(\ref{gw}).

As for the splitting of the decay width between the $1P_0$ and $1P_{\pm 1}$ states, based on the potential model in eq.~(\ref{iV1newmodel}), we find that
\be\la{spw2}
\Delta \Gamma= \Gamma(1P_0)-\Gamma(1P_{\pm 1})= -\frac{4}{5} \xi \lambda  c^\prime\frac{\alpha m_D^2- 4\sigma}{m_D^2}\, ,
\ee
Notice that the term $\sim \xi {\hat r}^2 \ln {\hat r}$ in the expansion in eq.~(\ref{iV1newmodel}) gives sub-leading contribution to $\Delta \Gamma$ which is neglected in the above equation. As compared to eq.~(\ref{spw}), the above splitting appears at a different order in the small ${\hat r}$ expansion. Since $\alpha m_D^2 < 4\sigma$ can be well satisfied, eq.~(\ref{spw2}) indicates that $1P_0$ state has a smaller decay width than $1P_{\pm 1}$ due to the negative $c^\prime$. Given the result in eq.~(\ref{spnew}), we can conclude that $1P$ states with $L_z=0$ have a higher dissociation temperature as compared to those with $L_z=\pm 1$. This is opposite to the result obtained by using the potential model discussed in sec.~\ref{apot}.

Although the same asymptotic limit when $r\rightarrow \infty$ can be observed in both anisotropic HQ potential models, only the one discussed in sec.~\ref{apot} shows the desired behavior as $r\rightarrow 0$. Furthermore, using the perturbative terms in eqs.~(\ref{rV1old}) and (\ref{iV1newmodel}) and setting ${\tilde m}_D/m_D=1-\xi(3+\cos (2\theta))/16$ and ${\hat \lambda}(\xi,\theta)/\lambda={\hat m}_D(\lambda, \xi,\theta)/m_D\approx1-\xi(0.175+0.026 \cos (2\theta))$ which corresponds to a matching at ${\hat r}=1$, a considerable discrepancy at small ${\hat r}$ is found when compared with the exact result. Therefore, it appears to be more reasonable to use eqs.~(\ref{rV1}) and (\ref{iV1}) to model the HQ potential in the anisotropic QGP. We emphasize that uncertainties in the non-perturbative part of the potential models, especially in the energy splitting between the p-wave states can not be settled at present. Although it seems very difficult, an experimental measurement on the polarization of $\chi_b(1P)$ state could provide a way to assess the correctness of these models.

\begin{table}[htpb]
\begin{center}
\setlength{\tabcolsep}{5mm}{
\begin{tabular}{  c  c  c  c c c }
\toprule[1pt]
$    \Upsilon(1S)     $  & ${\rm Re} E $  & $ \delta{\rm Re} E$ & $E_{\rm bind}$ &  $  {\rm Im} E  $ & $ \delta {\rm Im} E $ \\ \hline
$       T_o       $  & $182.869 $  &  $ 0.611  $ & $-662.669   $  &  $ 11.838     $ & $ 0.027 $   \\
$    1.1T_o       $  & $174.957 $  &  $ 0.593  $ & $-570.612   $  &  $ 14.830     $ & $ 0.031 $   \\
$    1.2T_o       $  & $166.556 $  &  $ 0.573  $ & $-493.689   $  &  $ 18.190     $ & $ 0.034 $   \\
$    1.4T_o       $  & $148.439 $  &  $ 0.531  $ & $-372.540   $  &  $ 26.004     $ & $ 0.039 $   \\
$    1.6T_o       $  & $128.807 $  &  $ 0.484  $ & $-281.672   $  &  $ 35.245     $ & $ 0.041 $   \\
$    1.8T_o       $  & $107.915 $  &  $ 0.435  $ & $-211.240   $  &  $ 45.833     $ & $ 0.040 $   \\
$    2.0T_o       $  & $85.978  $  &  $ 0.384  $ & $-155.279   $  &  $ 57.659     $ & $ 0.036 $   \\
$    2.1T_o       $  & $74.670  $  &  $ 0.359  $ & $-131.473   $  &  $ 63.998     $ & $ 0.033 $   \\
$    2.2T_o       $  & $63.160 $  &   $ 0.333  $ & $-109.961   $  &  $ 70.597     $ & $ 0.029 $   \\
\bottomrule[1pt]	
\end{tabular}
\vspace{0.2cm}
\vspace{0.2cm}
\begin{tabular}{  c  c  c  c c c}
\toprule[1pt]
$    \chi_{b0}(1P)      $  & ${\rm Re} E$  & $ \delta{\rm Re} E$ &  $E_{\rm bind}$  & $ {\rm Im} E$ & $ \delta {\rm Im} E$  \\ \hline
$       T_o         $  & $492.974$   & $1.444 $ & $-352.564   $  & $35.872$  & $0.132$  \\
$    1.1T_o         $  & $475.762$   & $1.345 $ & $-269.808   $  & $44.749$  & $0.131$  \\
$    1.2T_o         $  & $457.998$   & $1.246 $ & $-202.248   $  & $54.566$  & $0.123$  \\
$    1.3T_o         $  & $439.822$   & $1.149 $ & $-146.364   $  & $65.232$  & $0.107$  \\
$    1.4T_o         $  & $421.347$   & $1.057 $ & $-99.632    $  & $76.643$  & $0.085$  \\
\bottomrule[1pt]	
\end{tabular}
\vspace{0.2cm}
\vspace{0.2cm}
\begin{tabular}{  c  c  c  c c c}
\toprule[1pt]
$ \chi_{b\pm1}(1P)      $  & ${\rm Re} E$  & $ \delta{\rm Re} E$ &  $E_{\rm bind}$  & $ {\rm Im} E$ & $ \delta {\rm Im} E$  \\ \hline
$       T_o         $  & $461.960$   &$ 0.997       $ & $-383.578       $   &$ 34.996 $ &  $0.097$  \\
$    1.1T_o         $  & $446.761$   &$ 0.935       $ & $-298.809       $   &$ 43.448 $ &  $0.099$  \\
$    1.2T_o         $  & $431.014$   &$ 0.872       $ & $-229.231       $   &$ 52.752 $ &  $0.097$  \\
$    1.3T_o         $  & $414.833$   &$ 0.810       $ & $-171.353       $   &$ 62.831 $ &  $0.090$  \\
$    1.4T_o         $  & $398.307$   &$ 0.750       $ & $-122.672       $   &$ 73.599 $ &  $0.079$  \\
\bottomrule[1pt]	
\end{tabular}
\vspace{0.2cm}
\vspace{0.2cm}
\begin{tabular}{  c  c  c  c c c}
\toprule[1pt]
$      J/\Psi       $  & ${\rm Re} E$   & $ \delta{\rm Re} E$ &  $E_{\rm bind}$  & ${\rm Im} E$   & $ \delta{\rm Im} E $ \\ \hline
$       T_o         $  & $439.336 $   & $1.230 $ & $ -406.202 $   & $41.980$ & $ 0.107  $  \\
$    1.1T_o         $  & $422.207 $   & $1.163 $ & $ -323.362 $   & $51.467$ & $ 0.105  $  \\
$    1.2T_o         $  & $404.597 $   & $1.095 $ & $ -255.648 $   & $61.698$ & $ 0.098  $  \\
$    1.3T_o         $  & $386.604 $   & $1.028 $ & $ -199.583 $   & $72.564$ & $ 0.086  $  \\
$    1.4T_o         $  & $368.301 $   & $0.963 $ & $ -152.678 $   & $83.958$ & $ 0.070  $  \\
\bottomrule[1pt]	
\end{tabular}
}
\end{center}
\caption{The exact results of the complex eigenenergies ($E$) and binding energies ($E_{\rm bind}$) for different quarkonium states at various temperatures with $\xi=1$. Comparing with the results obtained based on the 1D potential model with effective screening masses, the corresponding differences as denoted by $\delta E$ are also listed. The reference temperature $T_o$ is $192\,{\rm {MeV}}$ and all the results are given in the units of ${\rm {MeV}}$.}
\label{xi1}
\end{table}

\begin{table}[htpb]
\begin{center}
\setlength{\tabcolsep}{10mm}{
\begin{tabular}{  c  c  c  c }
\toprule[1pt]
$    \Upsilon(1S)     $  & ${\rm Re} E $  & $ E_{\rm bind} $ & $  {\rm Im} E  $  \\ \hline
$       T_o       $  & $167.137 $  & $ -493.108   $ &  $ 18.157   $   \\
$    1.1T_o       $  & $156.439 $  & $ -416.062   $ &  $ 22.668  $   \\
$    1.2T_o       $  & $145.154 $  & $ -351.807   $ &  $ 27.702  $   \\
$    1.4T_o       $  & $121.059 $  & $ -250.972   $ &  $ 39.285  $   \\
$    1.6T_o       $  & $95.276  $  & $ -175.798   $ &  $ 52.753  $   \\
$    1.7T_o       $  & $81.859  $  & $ -145.082   $ &  $ 60.131  $   \\
$    1.8T_o       $  & $68.135  $  & $ -117.962   $ &  $ 67.900  $   \\
\bottomrule[1pt]	
\end{tabular}
\vspace{0.2cm}
\vspace{0.2cm}
\begin{tabular}{  c  c  c  c }
\toprule[1pt]
$    \chi_{b}(1P) $ & ${\rm Re} E $  & $ E_{\rm bind} $ & $  {\rm Im} E  $  \\ \hline
$       T_o       $  & $441.244 $  & $ -219.002   $ &  $ 53.216  $   \\
$    1.1T_o       $  & $420.901 $  & $ -151.600   $ &  $ 65.639  $   \\
$    1.2T_o       $  & $400.106 $  & $ -96.855    $ &  $ 79.060  $   \\
\bottomrule[1pt]	
\end{tabular}
\vspace{0.2cm}
\vspace{0.2cm}
\begin{tabular}{  c  c  c  c c c}
\toprule[1pt]
$      J/\Psi       $ & ${\rm Re} E $  & $ E_{\rm bind} $ & $  {\rm Im} E  $  \\ \hline
$       T_o       $  & $405.703 $  & $ -254.542   $ &  $ 61.604  $   \\
$    1.1T_o       $  & $383.991 $  & $ -188.510   $ &  $ 74.726  $   \\
$    1.2T_o       $  & $361.851 $  & $ -135.109   $ &  $ 88.583  $   \\
\bottomrule[1pt]	
\end{tabular}
}
\end{center}
\caption{The results of the complex eigenenergies ($E$) and binding energies ($E_{\rm bind}$) for different quarkonium states at various temperatures with $\xi=0$. The reference temperature $T_o$ is $192\,{\rm {MeV}}$ and all the results are given in the units of ${\rm {MeV}}$.}
\label{xi0}
\end{table}

\section{The eigen/binding energies and decay widths of the heavy-quarkonium bound states}\label{appa}

With the complex HQ potential model given in eqs.~(\ref{rV1}) and (\ref{iV1}) where the anisotropic screening masses are determined by eqs.~(\ref{aeffre}) and (\ref{aeffim}) for the real and imaginary part, respectively, we numerically solve the three-dimensional Schr{\"o}dinger equation by using a previously developed code called quantumFDTD~\cite{Strickland:2009ft,Delgado:2020ozh}. In Table~\ref{xi1}, we list the exact results of the eigenenergies (${\rm Re}\,E $), decay widths ($ {\rm Im}\,E$) as well as the binding energies ($E_{\rm bind}$) with the anisotropy parameter $\xi=1$ for several low-lying heavy-quarkonium bound states, including $\Upsilon(1S)$, $\chi_{b0}(1P)$, $\chi_{b\pm1}(1P)$, and $J/\Psi$. We consider various temperatures up to the dissociation temperature where the magnitude of the binding energy equals twice the decay width. Comparing with the results obtained using the 1D effective potential model with effective screening masses, the corresponding differences as denoted by ${\rm Re}\, \delta E$ and $ {\rm Im} \,\delta E$ are also listed for directly testing our method. In addition, the results in the isotropic limit with $\xi=0$ are provided in Table~\ref{xi0} for demonstrating the momentum-anisotropy effects.

In the numerical evaluations, we took $\alpha=0.272$ and  $\sigma=0.215\,{\rm GeV}^2$. For the $\Upsilon(1S)$ state, we used a lattice size of $N^3=512^3$ with a lattice spacing of $a=0.020\, {\rm GeV}^{-1}\approx 0.004 \,{\rm fm}$ giving a lattice size of $L=Na\approx 2.05\, {\rm fm}$. While for other three quarkonium states which have comparable root-mean-square radii, we used a lattice size of $N^3=256^3$ with a lattice spacing of $a=0.085\, {\rm GeV}^{-1}\approx 0.017\, {\rm fm}$ giving a lattice size of $L=Na\approx 4.35 \, {\rm fm}$. We have verified that, with the above lattice configurations, for real-valued isotropic potentials one can perfectly reproduce the eigenenergies obtained by using a one-dimensional Mathematica eigensolver (see, e.g., ref.~\cite{Lucha:1998xc}).

\section{Charmonium real-time evolution}
\label{app:charmonium}

\begin{figure}
\centering
\includegraphics[width=0.9\textwidth]{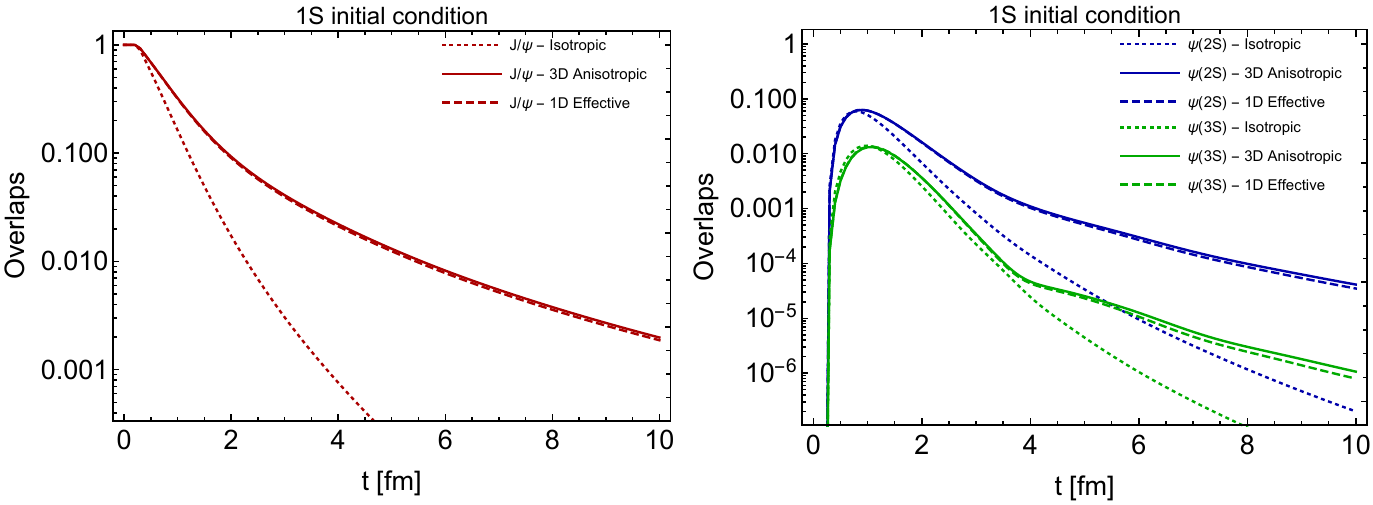}\\[1em]
\includegraphics[width=0.9\textwidth]{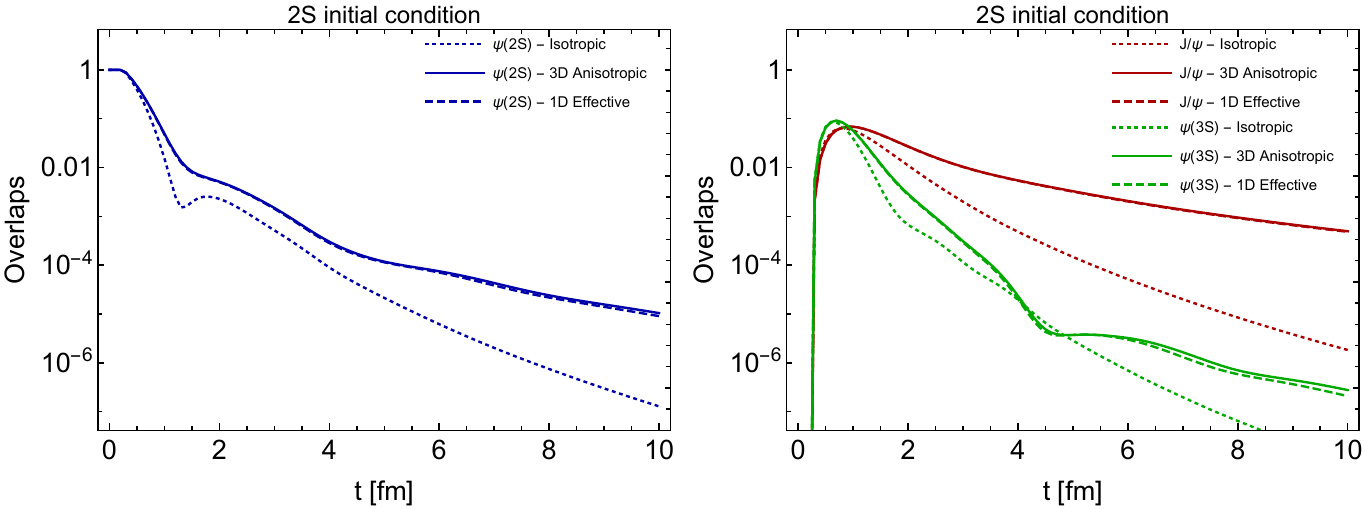}\\[1em]
\includegraphics[width=0.9\textwidth]{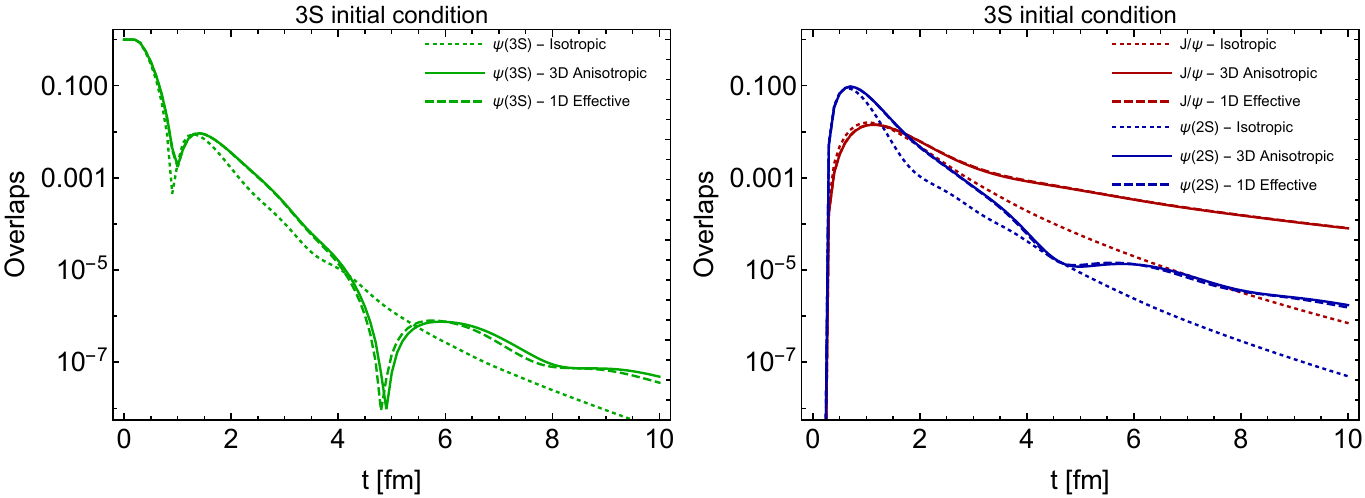}
\caption{Time evolution of the $J/\psi$, $\psi(2S)$, and $\psi(3S)$ overlaps.  The dotted, solid, and dashed lines correspond to using the 1D isotropic, 3D anisotropic, and 1D effective potentials, respectively.  The red, blue, and green colors correspond to the $J/\psi$, $\psi(2S)$, and $\psi(3S)$ overlaps, respectively.}
\label{fig:overlaps-jpsi}
\end{figure}

In this appendix we present results of the real time evolution of the vacuum overlaps for charmonium states obtained using the 1D effective potential and the full 3D anisotropic potential.  For charmonium states we take $L = $ 5.12 fm, $m_c = 1.3$ GeV, and use $N=128$ lattice points in each direction.  We use the same temporal lattice spacing $\Delta t$ as in the case of bottomonium.  As with bottomonium, we solve for the vacuum eigenstates and then either initialize with pure eigenstates or a Gaussian form.  In fig.~\ref{fig:overlaps-jpsi}, we present the time evolution of the $J/\psi$, $\psi(2S)$, and $\psi(3S)$ vacuum overlaps obtained with eigenstate initial conditions.  The dotted, solid, and dashed lines correspond to using the 1D isotropic, 3D anisotropic, and 1D effective potentials, respectively.  The red, blue, and green colors correspond to the $J/\psi$, $\psi(2S)$, and $\psi(3S)$ overlaps, respectively.  As can be seen from this figure, similar to bottomonium, the one dimensional effective potential model can accurately describe the results obtained with the anisotropic three-dimensional potential, while the isotropic model fails to describe the full 3D evolution well.  We see slightly larger differences between the 1D effective and 3D evolutions for charmonium than seen for bottomonium due to the fact that the one-dimensional effective potential model is optimized for small to medium $r$ and hence states with small $\langle r \rangle$; however, the method also works quite well for the charmonium overlap evolution despite this.

\begin{figure}
\centering
\includegraphics[width=1\textwidth]{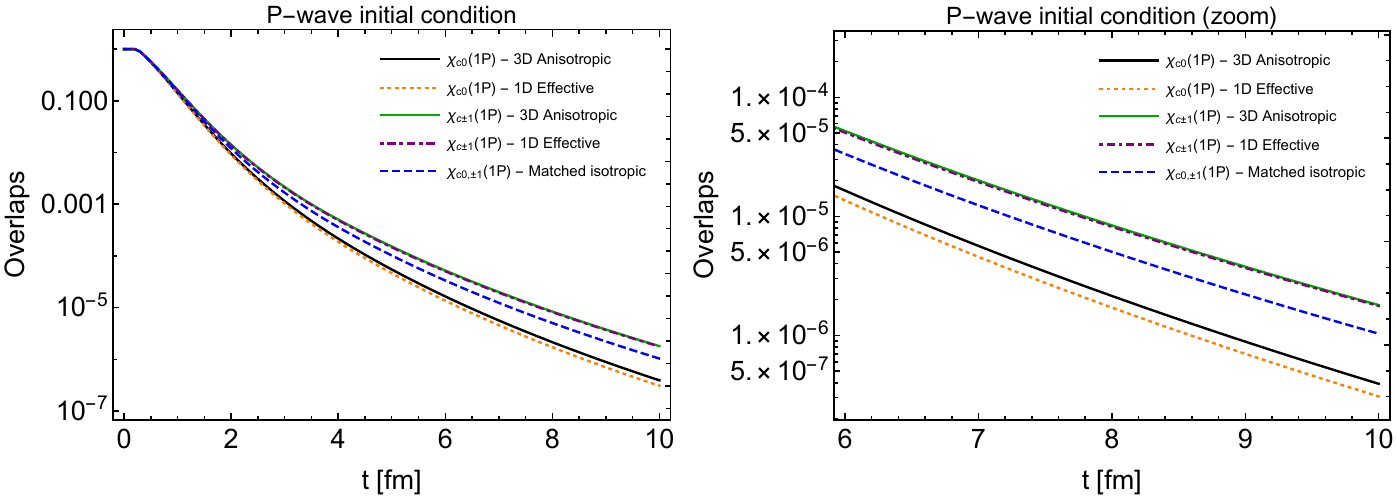}
\caption{Time evolution of the charmonium p-wave overlaps resulting from initialization with different p-wave polarizations corresponding to $l=1$ and $m=0,\pm1$ labeled as $\chi_{c0}(1P)$ and $\chi_{c\pm1}(1P)$, respectively.  The solid black and orange lines correspond to the full 3D evolution with $\chi_{c0}(1P)$ and $\chi_{c\pm1}(1P)$ initial conditions and the orange dotted and purple dot dashed lines correspond to the 1D effective potential evolution with the same initial conditions. The dashed blue line corresponds to the isotropic matching scheme detailed in the main body of the text.}
\label{fig:overlaps-pwave-jpsi}
\end{figure}

\begin{figure}
\centering
\includegraphics[width=0.8\textwidth]{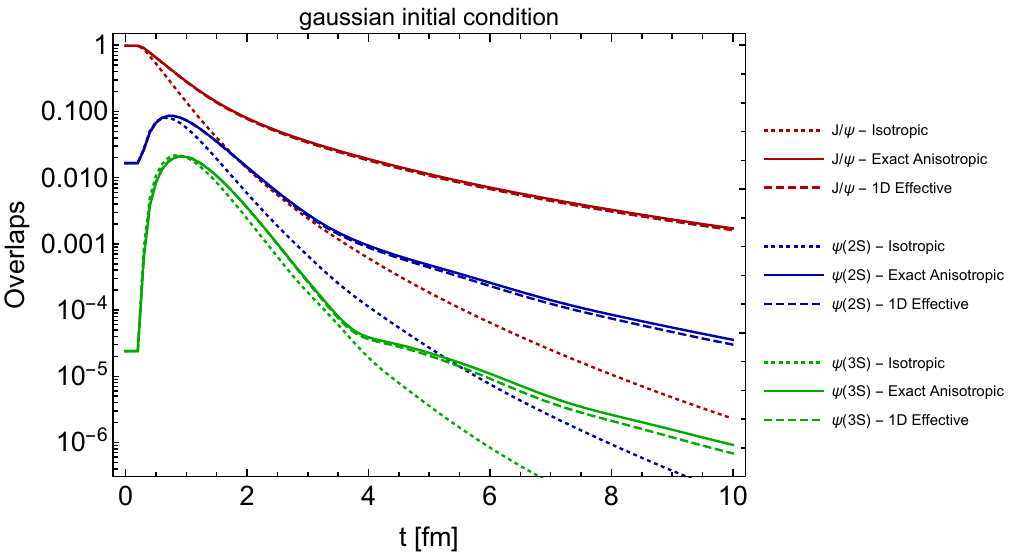}
\caption{Time evolution of the $J/\psi$, $\psi(2S)$, and $\psi(3S)$ overlaps.  The dotted, solid, and dashed lines correspond to using the 1D isotropic, 3D anisotropic, and 1D effective potentials, respectively.  The red, blue, and green colors correspond to the $J/\psi$, $\psi(2S)$, and $\psi(3S)$ overlaps, respectively.
}
\label{fig:overlaps-gaussian-jpsi}
\end{figure}

Turning to p-wave initial conditions, in fig.~\ref{fig:overlaps-pwave-jpsi} we present the time evolution of the charmonium p-wave overlaps resulting from initialization with different p-wave polarizations corresponding to $l=1$ and $m=0,\pm1$ labeled as $\chi_{c0}(1P)$ and $\chi_{c\pm1}(1P)$, respectively.  The solid black and orange lines correspond to the full 3D evolution with $\chi_{c0}(1P)$ and $\chi_{c\pm1}(1P)$ initial conditions and the orange dotted and purple dot dashed lines correspond to the 1D effective potential evolution with the same initial conditions. The dashed blue line corresponds to the isotropic matching scheme detailed in the main body of the text.  This figure demonstrates that, similar to bottomonium, the 1D effective model can well describe the time evolution of the splitting between different p-wave polarizations.  We note that due to the lower mass of the charmonium states, the observed p-wave splitting is larger and, as a result, the matched isotropic approximation does more poorly in describing the time evolution of the vacuum overlaps than in the case of bottomonium.

Finally, in fig.~\ref{fig:overlaps-gaussian-jpsi} we present a comparison of the 1D and 3D real-time evolution obtained using a Gaussian initial condition of the form given in eq.~\eqref{eq:gaussianform}.  In this figure the dotted, solid, and dashed lines correspond to using the 1D isotropic, 3D anisotropic, and 1D effective potentials, respectively.  The red, blue, and green colors correspond to the $J/\psi$, $\psi(2S)$, and $\psi(3S)$ overlaps, respectively.  As can be seen from this figure, the 1D effective potential model once again provides an excellent approximation to the full 3D potential evolution, while the isotropic model fails to describe the full 3D evolution.


\end{document}